\documentclass{aa}

\usepackage{graphicx}
\usepackage{caption}
\usepackage{subcaption}
\usepackage{txfonts}
\usepackage{xcolor}
\usepackage[breaklinks,colorlinks,urlcolor=blue,citecolor=blue,linkcolor=blue]{hyperref}

\begin{document}

   \title{Predicted white-light solar flare emission from the F-CHROMA grid of models}

   \author{S. Ornig\inst{1} \fnmsep \inst{2}\and
          M. Carlsson\inst{1} \fnmsep \inst{2}}

   \institute{Institute of Theoretical Astrophysics, University of Oslo,
              PO Box 1029 Blindern, 0315 Oslo, Norway\\
              \email{sascha.ornig@astro.uio.no}
         \and
             Rosseland Centre for Solar Physics, University of Oslo,
              PO Box 1029 Blindern, 0315 Oslo, Norway\\
             }

   \date{Received dd.mm.yyyy; accepted dd.mm.yyyy}

  \abstract
   {Much of a solar flare's energy is thought to be released in the continuum. The optical continuum (white light) is of special interest due to the ability to observe it from the ground.}
   {We aim to investigate the prevalence of white-light (WL) emissions in simulations of purely electron beam-driven solar flares, what determines the occurrence of these enhancements, and the underlying causes.}
   {We utilized the F-CHROMA grid of flare simulations created using the radiative hydrodynamics code \mbox{RADYN}. We probed the spectral index, total energy, and low-energy cutoff to draw conclusions about their relationships to the white-light intensity. Furthermore, we calculated the 6684~\AA~continuum intensities, the Balmer, and the Paschen ratios. Finally, we analyzed two particular cases, one with high 6684~\AA~intensity and one with a large Balmer ratio, to determine the dominant mechanisms in these simulations.}
   {33 of the 84 flares included in the F-CHROMA grid show white-light intensity enhancements that exceed 0.1\% relative to the pre-flare level. We conclude that, with the parameters presented in the F-CHROMA grid, purely electron beam-driven simulations of solar flares are not able to reproduce observed WL enhancements, as the maximum enhancements in the grid are below 4\%, which is significantly lower than observational values. The total energy (which is correlated with the maximum beam flux) is the main factor for deciding whether excess white-light emissions will be detectable or not. There is a linear relationship between the Balmer (and Paschen) ratio and the relative continuum increase. Both case studies show that during the time of maximum WL excess, hydrogen ionization and subsequent recombination in an optically thin medium is the dominant mechanism for WL continuum emission enhancements in these electron beam-driven atmospheres. Increased H$^-$~emission in the photosphere as a result of radiative backwarming becomes dominant during the declining phase of WL emissions in both case studies. We confirm the inability to reproduce the characteristics of type II WL flares, namely the appearance of photospheric continuum enhancements without prior chromospheric emission increases and the lack of a clear Balmer jump, with the F-CHROMA grid.}
   {}


   \maketitle


\section{Introduction}


Flares are sudden local brightenings that appear throughout the electromagnetic (EM) spectrum over a timespan of minutes to hours \citep{2017LRSP...14....2B}. They can accelerate particles to high energies, heat the ambient plasma, and be accompanied by coronal mass ejections \citep{2011LRSP....8....6S}. To explain the observations, the so-called \mbox{CSHKP} model (or Standard Flare Model; named after \citealt{1964NASSP..50..451C}, \citealt{1966Natur.211..695S}, \citealt{1974SoPh...34..323H}, \citealt{1976SoPh...50...85K}) was developed. A large magnetic loop rises up to coronal heights due to an instability in the system and gets pinched at its legs. Where the anti-parallel magnetic field lines meet, magnetic reconnection sets in and releases the built-up energy. Charged particles (electrons and ions) are accelerated and move along the magnetic loop toward the chromosphere. There, the density is high enough that the energetic particles can lose their energy through Coulomb collisions, thereby heating the ambient plasma. Since the plasma pressure is dependent on the temperature, the heated plasma expands and fills the magnetic loop in a process known as chromospheric evaporation \citep{1985ApJ...289..414F}. Along the way, it emits radiation in the soft X-ray (SXR) and extreme ultraviolet (EUV) parts of the EM spectrum \citep{2017LRSP...14....2B}.\\
\indent The first observations of solar flares were conducted in the optical part of the EM spectrum \citep{1859MNRAS..20...13C}. Here, especially the Balmer and Ca II K lines have become increasingly popular \citep{2011A&A...530A..84K}. However, spectral lines are not the only parts of the spectrum that contribute to optical emissions. The continuum has arguably received less attention due to the short duration and poor contrast of its intensities \citep{2006SoPh..234...79H, 2007ApJ...656.1187F}. It was postulated that emissions in the optical continuum, known as white-light (WL) emissions, not only occur in certain conditions, but rather could be general features of flares \citep{2008ApJ...688L.119J}. In a statistical analysis, \cite{2017ApJ...850..204W} found WL emission in half of a sample of 100~M- and X-class flares observed by Hinode/SOT. Another, more recent study by \cite{2024ApJ...975...69C} randomly selected a set of flares from the flare list in \cite{2021ApJ...917L..29L} (and added 6 X-class flares on their own). Using a revised method of identification applied to SDO/HMI continuum observations, they found WL emission in 30\% of the 30~C-class flares they studied, with an even higher proportion for M- and X-class flares. Similar values have been reported by \cite{2018A&A...613A..69S} using observations taken from space (by SDO/HMI) and the ground (by the Optical and Near-infrared Solar Eruption Tracer, ONSET), although the differentiation between C- and M-class flares was not considered. \cite{2020ApJ...904...96C} report comparable percentages for M- and X-class flares, but lower (\mbox{$\approx 10$\%}) for C-class events; results that match those for circular-ribbon flares in \cite{2018ApJ...867..159S}. A concise summary of statistics on WL flare occurrences across several studies can be found in \cite{2020ApJ...904...96C}.\\
\indent In WL flares, the optical continuum contains a significant portion of the total radiated energy and may even play the dominant role overall. \cite{1989SoPh..121..261N} suggested that WL intensity enhancements may account for up to 90\% of the total energy emitted from the flare region. This is supported by findings in Sun-as-a-star observations by \cite{2011A&A...530A..84K}, who found that about two-thirds of the radiated flare energy is contained in WL. On the other hand, more recent statistical investigations found that the relative enhancement of the optical continuum at 360~nm during WL flares is less than 30\% in most cases \citep{2024SoPh..299...11J}, with a mean of approximately 19\%. This seems to be strongly dependent on the resolution of the instrument, as \cite{2008ApJ...688L.119J} found enhancements of up to 300\% in localized regions.\\
\indent As for the origin of WL emissions, the mechanisms are still poorly understood. The theories can be divided mainly into two categories \citep{1994ApJ...429..890D}: The first, where the emission comes from regions that emit at the same wavelengths in the quiet Sun, i.e.,~the photosphere \citep{2014ApJ...783...98K}, and the second, where the emission comes from higher up, i.e.,~the chromosphere. In the former, the energy needs to be transported to the photosphere by high-energy electron beams \citep{1978ApJ...224..241E, 2018ApJ...862...76P}, Alfvén waves \citep{2013ApJ...765...81R} or radiative backwarming \citep{1989SoPh..124..303M, 2003ApJ...595..483M}. In the latter, WL emission is characterized by optically thin recombination radiation, mainly in hydrogen continua \citep{2018ApJ...867L..24D}. A more detailed discussion of possible WL emission mechanisms can be found in \cite{1990ApJ...365..391M}.\\
\indent Following these two theories, WL flares can then be broadly classified into two groups depending on their spectral properties, as well as the relation between intensity enhancements of the WL and that of hard X-rays (HXR) or microwaves \citep{1986A&A...159...33M}. Type~I~WL~flare spectra show a Balmer jump and strong Balmer line intensities. WL emission of type~I~WL~flares generally coincides in time with HXR or microwave enhancements \citep{1995A&AS..110...99F,2016ApJ...816....6K}, and is believed to be caused by hydrogen recombination in the chromosphere \citep{2016ApJ...816...88K}. Type~II~WL~flares, on the other hand, lack a clear Balmer jump, show weak and narrow hydrogen lines, and precede or lag behind HXR and microwave increases \citep{1995A&AS..110...99F,1999ApJ...512..454D}. The mechanism behind type~II~WL~flares is thought to be a higher H$^-$~continuum emission from the photosphere \citep{2016ApJ...816...88K}.\\
\indent In this study, we aim at investigating the mechanisms responsible for white-light enhancements in 1D purely electron beam-driven simulations of solar flares. To accomplish this, we made use of the F-CHROMA grid presented by \cite{2023A&A...673A.150C}, which includes a range of simulations with electron beam parameters realistic for solar conditions.


\section{Methodology}


\subsection{The radiative hydrodynamics code RADYN}

To simulate flares and the response of the solar atmosphere, we need to combine the radiative transfer equation with the hydrodynamic equations. This necessitates a model that extends from below the photosphere through the chromosphere, transition region, and finally the corona. The \mbox{RADYN} code, developed by \cite{1992ApJ...397L..59C, 1995ApJ...440L..29C, 1997ApJ...481..500C, 2002ApJ...572..626C} and \cite{2015ApJ...809..104A}, employs a semi-implicit (multidimensional) Newton--Raphson linearization and iterates the time-centered spatial derivatives on an adaptive grid \citep[see][]{1987JCoPh..69..175D}. In extreme instances, the minimum distance between two grid points may reach less than 1 m. To model transitions, \mbox{RADYN} incorporates a five-level-plus-continuum hydrogen atom, an eight-level-plus-continuum helium atom, and a five-level-plus-continuum calcium atom. The continua of other elements are treated as background continua in LTE \citep{2002ApJ...572..626C}.\\
\indent \cite{2023A&A...673A.150C} constructed a grid of flare simulations (the F-CHROMA grid), which is freely available\footnote{Models available at \url{https://star.pst.qub.ac.uk/wiki/public/solarmodels/start.html}}$^,$\footnote{RADYN source code available at \url{https://folk.universitetetioslo.no/matsc/radyn/}} and includes input files for the grid and a script to conduct the flare simulation. The atmosphere consists of a loop with a half-length of 10~Mm and a quarter-circle shape. The loop is assumed to be vertical, with the bottom boundary being situated 90~km below the \mbox{$\tau_{500nm} = 1$} height. Both the lower and upper boundaries are closed (i.e., reflective). The electrons in the beam are assumed to be non-thermal, having been accelerated in the corona, and to follow a power-law distribution. The shape of the electron beam flux in this model grid is triangular in time; it increases linearly for the first half (\mbox{$t = 10$}~s) and decreases linearly back to 0 during the second half. The beam follows the Fokker--Planck description \citep[incorporated by][]{2015ApJ...809..104A}. The characteristic properties of a given beam are the spectral index, $d$, the cutoff energy (the minimum non-thermal electron energy), $E_c$, and the total energy per area, $E_{tot}$. In the present stage, return currents (which provide additional heating) are not included in these models. To capture the development of the atmosphere after direct beam heating ceases, the simulations were extended for a further 30~s until \mbox{$t = 50$}~s.


\subsection{Parameters}

The respective parameters included in our analysis are:
\begin{enumerate}
    \item The spectral index, $d$, total energy, $E_{tot}$, and cutoff energy, $E_c$. We therefore specify simulations according to the following nomenclature: $\mathrm{dn\_E_{tot}\_E_c}$, where $n$ is a positive integer stating the spectral index. The units of $E_{tot}$ and $E_c$ are erg~cm$^{-2}$ and keV, respectively. For example, $\mathrm{d4\_1.0e12\_25}$ describes a simulation with spectral index $d = 4$, total energy $E_{tot} = 1\cdot10^{12}$~erg~cm$^{-2}$, and cutoff energy $E_c = 25$~keV. In total, the spectral index in the F-CHROMA grid ranges from \mbox{$d = 3$} to \mbox{$d = 8$}, which are reasonably common values found for solar flares \citep{1976SoPh...50..153L,2006ApJ...641.1217C,2007ApJ...656.1187F,2009ApJ...699..968M,2016A&A...588A.115W,2017ApJ...850..204W,2018ApJ...867..159S,2023ApJ...943L...6W}. The values for the total energy of the beam are \mbox{$E_{tot} = [3\cdot10^{10}$}, \mbox{$10^{11}$}, \mbox{$3\cdot10^{11}$}, \mbox{$10^{12}]$~erg~cm$^{-2}$}, with the higher end being larger than the values found in \cite{2016A&A...588A.115W}, but still below the highest values obtained from solar flares \citep{2024ApJ...969..121K}. The cutoff energies in the grid are \mbox{$E_c = $}~[10, 15, 20, 25, 38]~keV, which is consistent with typical values reported in the literature \citep{2006ApJ...641.1217C,2007ASPC..368..423F,2009ApJ...699..968M,2016A&A...588A.115W}. However, there is only one run with \mbox{$E_c = 38$}~keV. In theory, this assortment of parameters would result in a total of \mbox{$6 \cdot 4 \cdot 4 + 1 = 97$} simulations. In practice, some of the simulations with high $d$, high $E_{tot}$, and low $E_c$ are not included in the grid due to the enormous computational effort needed to simulate these events. At the time of writing, the F-CHROMA grid contains a total of 84 flare simulations.
    \item The continuum intensity, $I_c$, and its relative change (in \%), $I_{c,rel}$, compared to the $t = 0$ (pre-flare) value. We denote maximum continuum intensity for each simulation as $I_{c,max}$.\footnote{We will call this the WL maximum from here on.} The selected wavelength to represent the white-light continuum is 6684~\AA, which is the closest available wavelength to the Fe~I~6173~\AA~line used for the reconstruction of the SDO/HMI continuum \citep{2012SoPh..275..207S}.
    \item The Balmer jump ratio, $R_B$, which is the intensity blueward of the Balmer jump divided by the intensity redward. This is a direct result of the structure of the H-atom. Blueward of the Balmer jump (i.e., toward higher energies), hydrogen can be ionized directly from the second energy level due to bound-free absorption. The calculation of the Balmer ratio differs from that in \cite{2024ApJ...969..121K}, as we used the two wavelengths closest to the Balmer jump in the \mbox{RADYN} simulations: 3646.9~\AA~and 3647.1~\AA. In reality, the sharp edge that appears in simulations does not exist due to the blending of higher-order Balmer lines \citep[e.g.,][]{1985A&A...148..240Z}. Nonetheless, we calculated the absolute Balmer ratio as follows:
    \begin{equation}
        \hspace{0.2\textwidth} R_B = \frac{I_\mathrm{b,B}}{I_\mathrm{r,B}},
    \end{equation}
    with $I_\mathrm{b,B}$ the intensity blueward and $I_\mathrm{r,B}$ the intensity redward of the Balmer jump. We denote $R_B$ at the time of $I_{c,max}$ as $R_{B,max}$.
    \item The Paschen jump ratio, $R_P$, which originates from a similar configuration as the Balmer jump. However, in this case, the electron in the neutral hydrogen is already in state \mbox{$n = 3$}, and therefore the needed energy to ionize H is less than for \mbox{$n = 2$}. This results in the Paschen jump being situated at longer wavelengths. The Paschen ratio was analyzed using the intensities immediately blueward (8205.7~\AA) and redward (8206.0~\AA) of the jump, and the ratio was calculated similarly to the Balmer jump ratio:
    \begin{equation}
        \hspace{0.2\textwidth} R_P = \frac{I_\mathrm{b,P}}{I_\mathrm{r,P}},
    \end{equation}
    with $I_\mathrm{b,P}$ the intensity blueward and $I_\mathrm{r,P}$ the intensity redward of the Paschen jump. We denote $R_P$ at the time of $I_{c,max}$ as $R_{P,max}$.
    \item The temperature, density, velocity, and several other parameters related to the internal energy.
    \item The contribution function to intensity, which can be used to analyze where in the atmosphere the excess white-light emissions are coming from. This is a valuable analytical tool because it indicates the contributions from a certain depth point in the model to the total intensity at a given wavelength. Starting with the formal solution of the transfer equation \citep{1994chdy.conf...47C},
    \begin{equation}
         \hspace{0.155\textwidth} I_\nu = \int_{0}^{\infty} S_\nu \tau_\nu e^{-\tau_\nu} d\tau_\nu,
    \end{equation}
    we can identify the contribution function to the emergent intensity as a combination of the source function, $S_\nu$, an exponential attenuation factor, $e^{-\tau_\nu}$, and a term, $d\tau_\nu$, that describes the product of the cross-section and the column density of emitters. This can be rewritten as
    \begin{equation}\label{eq:contrib_func}
         \hspace{0.13\textwidth} I_\nu = \int_{-\infty}^{\infty} S_\nu(z) e^{-\tau_\nu} \chi_\nu(z) dz,
    \end{equation}
    where $\chi_\nu$ is the monochromatic opacity per volume. The term $S_\nu(z) e^{-\tau_\nu} \chi_\nu(z)$ in Eq.~\ref{eq:contrib_func} is the contribution function in terms of the geometrical height, z.
\end{enumerate}
\begin{figure*}
    \resizebox{\hsize}{!}
    {\includegraphics[width=0.49\textwidth]{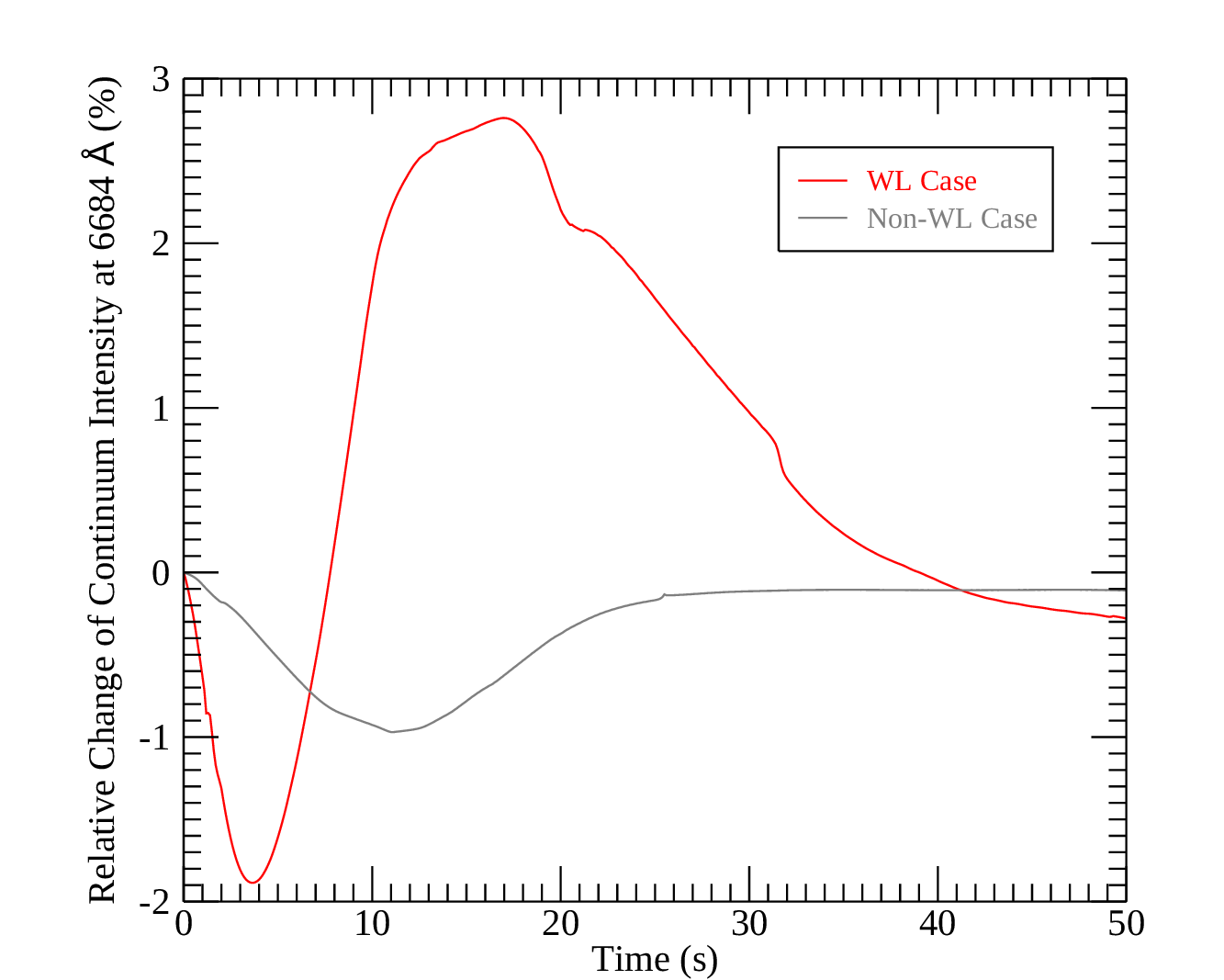}
    \includegraphics[width=0.49\textwidth]{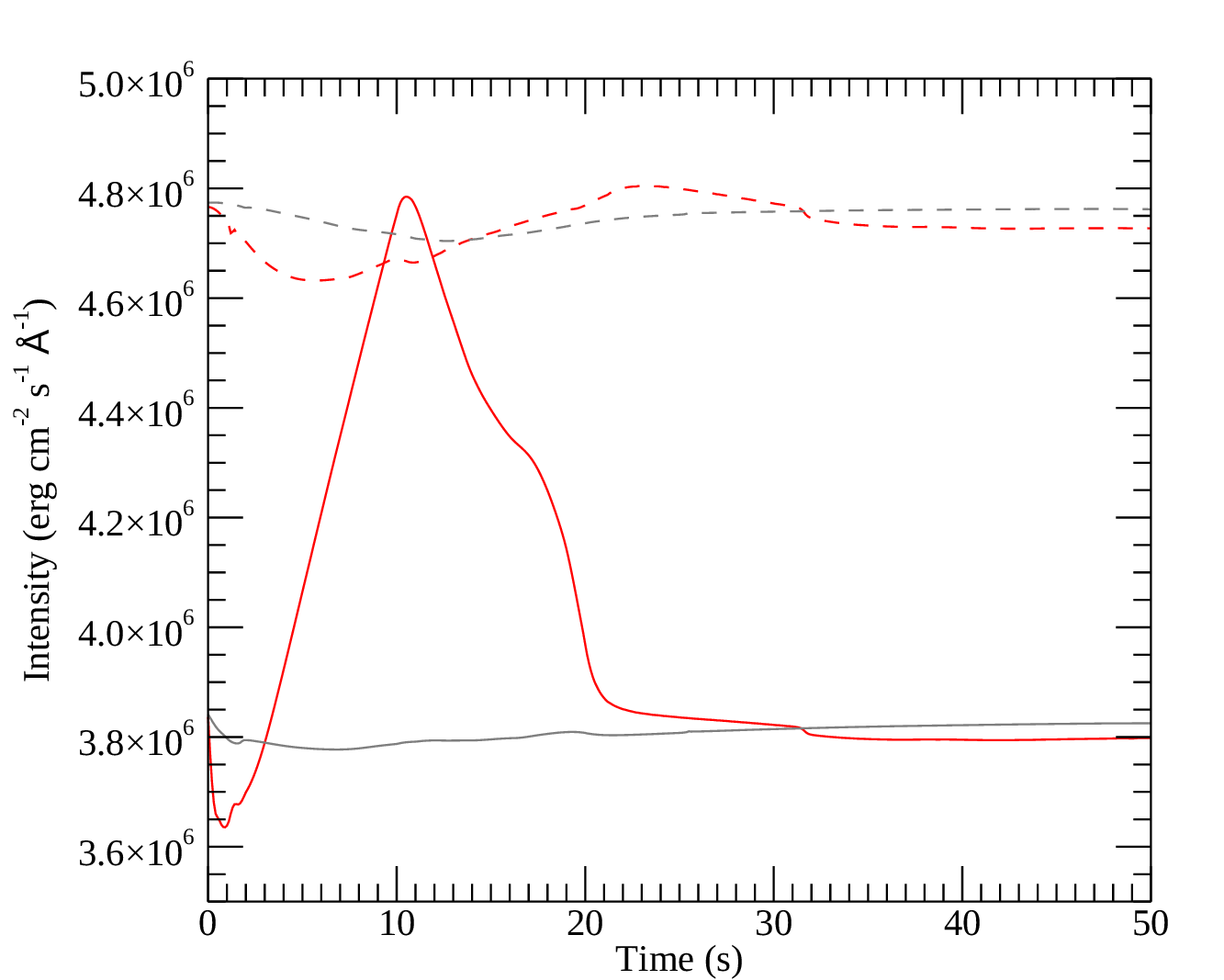}}
    \caption{Light curves of $I_{c,rel}$, as well as intensities left and right of the Balmer jump for a WL and a non-WL case. Left: $I_{c,rel}$ as a function of time for the WL case $\mathrm{d3\_1.0e12\_20}$ and the non-WL case $\mathrm{d3\_1.0e11\_10}$. Right: Intensities blueward (at 3646.9~\AA; solid lines) and redward (at 3647.1~\AA; dashed lines) of the Balmer jump. In both panels, red indicates the WL case, whereas gray specifies the non-WL case.}
    \label{fig:balmer_cont}
\end{figure*}


\section{Results} \label{sec:results}


\subsection{Statistical behavior} \label{subsec:statistics}

45 of the 84 cases show at least some enhancements of the continuum intensity at 6684~\AA, leaving 39 cases that do not produce any enhanced WL intensity. In fact, most of the cases (including the ones showing an increase in WL intensity) show some kind of decrease, in extreme cases down to over -2.0\%, usually reaching the peak decrease within the first 10~seconds, i.e.,~when the beam flux is still increasing. Fig.~\ref{fig:balmer_cont} displays $I_{c,rel}$ (left panel) as well as the Balmer jump light curves (blue- and redward; right panel) for two cases. One can see that the initial continuum decrease is steeper, but shorter in the case with WL intensity enhancements. This is in line with the results presented by \cite{2021RAA....21....1Y}, who propose that higher electron beam fluxes result in shorter initial dimming periods and that the severity of the dimming is inversely correlated with its duration.\\
\indent The events for Fig.~\ref{fig:balmer_cont} were chosen in order to demonstrate how in a non-WL case, the intensities around the Balmer jump generally stay approximately constant, except for small bumps and/or decreases, whereas in a WL case, there is usually a highly significant peak in the intensities blueward of the Balmer jump.\\
\indent Of the 45 flares that show $I_{c,rel} > 0$ at some point in time, only 33 have relative intensity enhancements that exceed 0.1\%.\footnote{From now on, we will use the term ``WL case'' for simulations showing a WL intensity enhancement over 0.1\% and ``non-WL case'' for the other simulations.} Only 12 flares exceed 1\% on this scale.\\
\indent A calculation of the portion of energy radiated in the continuum compared to the lines shows that on average, 46.6\% of the total energy radiated across the simulations included in the F-CHROMA grid is radiated in the continuum. For the non-WL cases, the average amount is 42.9\%, with values ranging between 21.2\% and 61.0\%. The respective values for WL cases range from 38.5\% to 73.8\%, with a mean of 52.5\%.\\
\begin{figure}
    \includegraphics[width=0.49\textwidth]{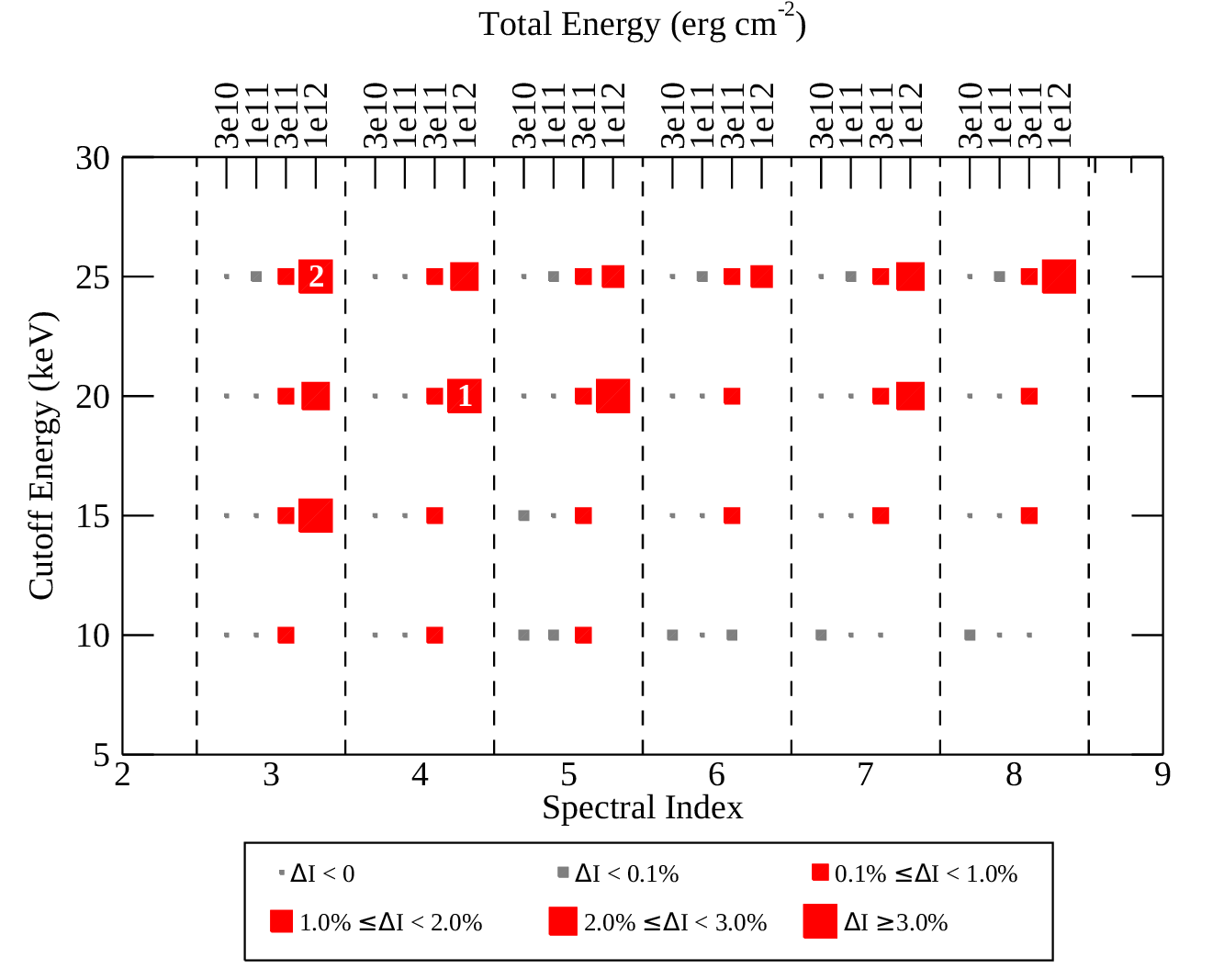}
    \caption{$I_{c,max}$ as a function of $E_c$, $d$, and $E_{tot}$ for all simulations included in the F-CHROMA grid. The values for $d$ have been divided into four distinct locations to avoid overlap and to show $E_{tot}$ corresponding to each simulation (indicated at the top). Dashed vertical lines indicate the border between the different spectral index values. The size corresponds to $I_{c,max}$, as indicated in the legend below the x-axis. Red indicates WL cases, whereas gray specifies non-WL cases. The cases used for the case studies are indicated with numbers 1 and 2.}
    \label{fig:fchroma_params}
\end{figure}
\indent Fig.~\ref{fig:fchroma_params} shows the maximum of $I_{c,rel}$ as a function of $d$, $E_{tot}$, and $E_c$.\footnote{The simulation with \mbox{$E_c = 38$}~keV is not included in the figure, but it produced a white-light enhancement of roughly 2.1\%.} The cases investigated in more detail in this work (Sect.~\ref{sec:event_1} and Sect.~\ref{sec:event_2}) are marked with numbers 1 and 2. The most dominant factor in determining whether a flare will have enhanced continuum intensities is the total energy (or maximum beam flux) of the electron beam. As depicted, atmospheres reacting to a total beam energy of \mbox{$E_{tot} \leq 1 \cdot 10^{11}$~erg~cm$^{-2}$} (i.e., a maximum beam flux of \mbox{$1 \cdot 10^{10}$~erg~cm$^{-2}$~s$^{-1}$} or lower) do not produce continuum intensity enhancements exceeding 0.1\% above the quiet-Sun level. From Fig.~\ref{fig:fchroma_params} it finally follows that WL emission enhancements are less likely for electron beam spectra with a combination of low $E_c$ and high $d$.\\
\begin{figure}
    \centering
    {\includegraphics[width=0.42\textwidth]{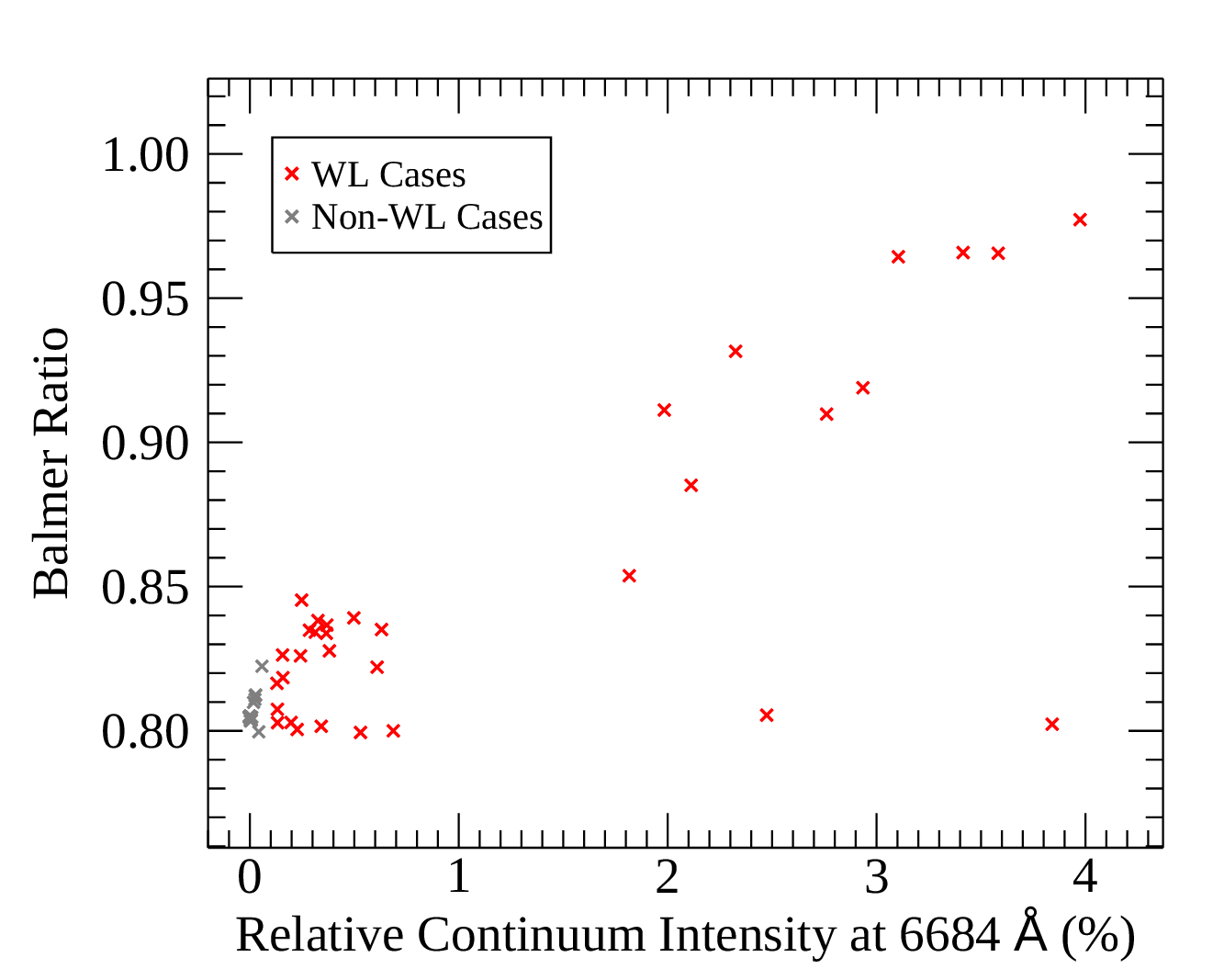}}
    {\includegraphics[width=0.42\textwidth]{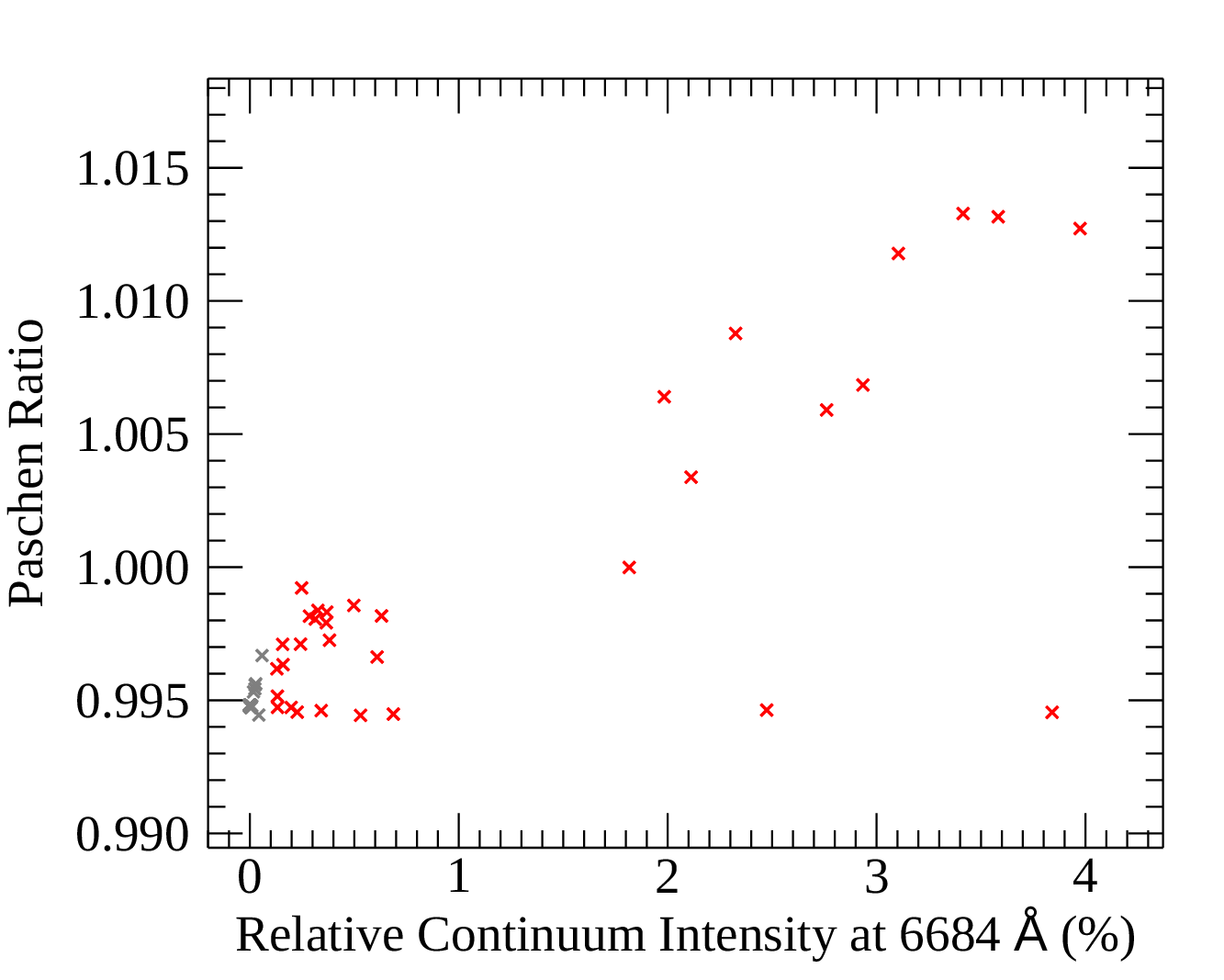}}
    {\includegraphics[width=0.42\textwidth]{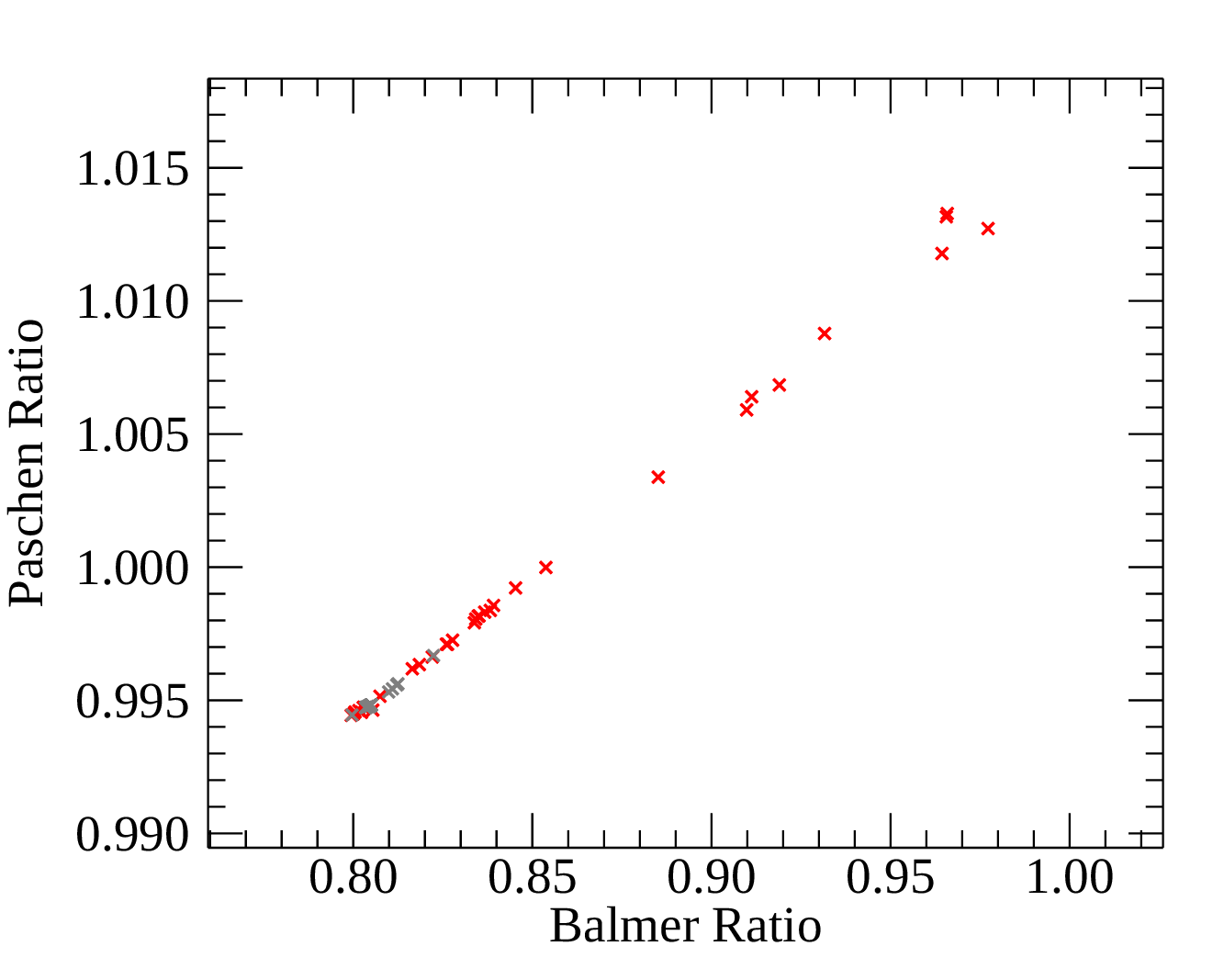}}
    \caption{Scatter plots of the $R_{B,max}$ (top panel) and $R_{P,max}$ (middle panel) as a function of $I_{c,max}$. The bottom panel shows $R_{P,max}$ as a function of $R_{B,max}$. In all panels, red indicates WL cases, whereas gray specifies non-WL cases.}
    \label{fig:ratio_scatters}
\end{figure}
\indent Fig.~\ref{fig:ratio_scatters} depicts the relationship between $R_{B,max}$ and $I_{c,max}$ (top panel), as well as between the latter and $R_{P,max}$ (middle panel). We note here that $R_{B,max}$ and $R_{P,max}$ are given at the time of $I_{c,max}$, and may therefore not reflect the maxima of the ratios for a given simulation. Interestingly, there seems to exist a linear relationship between $R_{B,max}$ (and $R_{P,max}$) and $I_{c,max}$. One of the two simulations in the bottom right of the upper two panels of Fig.~\ref{fig:ratio_scatters} shows peculiar behavior that is likely related to numerical issues. $R_{B,max}$ ranges from 0.80~to~0.98, whereas $R_{P,max}$ extends from 0.995 to 1.013. For reference, the value of $R_B$ at \mbox{$t = 0$} is 0.80, and 0.995 for $R_P$. This means that all cases with $R_{B,max} = 0.80$ and $R_{P,max} = 0.995$ did not exhibit an increase before $I_{c,rel}$ reached its maximum. Generally, cases with $R_{B,max} > 0.87$ all show WL enhancements of 0.1\% or more. However, several WL cases also show $R_{B,max} < 0.87$. $R_{P,max}$ does not show the same spread in values as $R_{B,max}$ does, but still a rough differentiation between non-WL and WL cases can be made, as all cases with $R_{P,max} > 0.997$ show $I_{c,max} > 0.1$\%. The bottom panel of Fig.~\ref{fig:ratio_scatters} presents the relationship between $R_{B,max}$ and $R_{P,max}$. The two exhibit very similar behavior, with an almost perfect linear correlation.


\subsection{Case~1: High continuum intensity enhancement}\label{sec:event_1}

\begin{figure}
    {\includegraphics[width=0.49\textwidth]{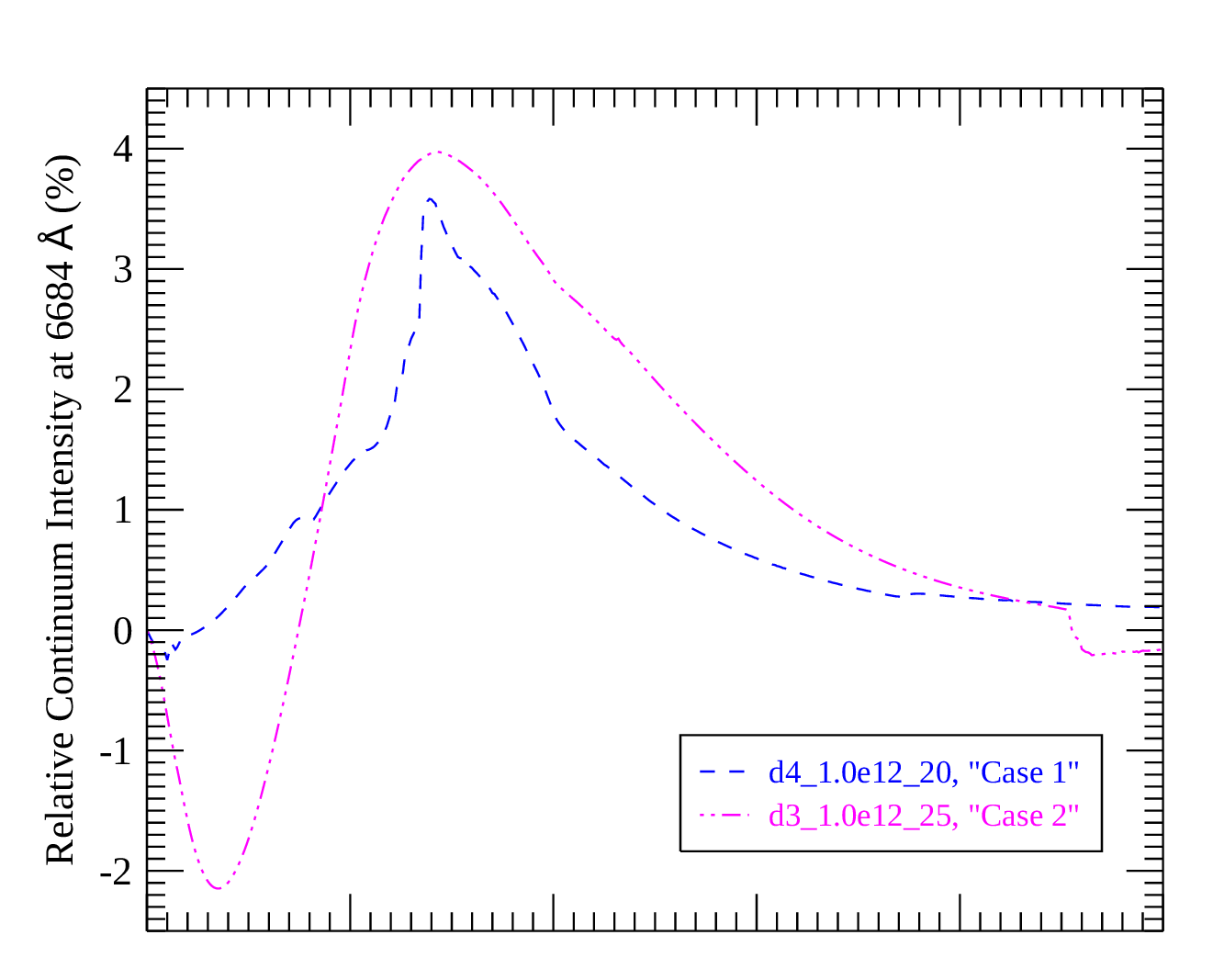}}
    {\includegraphics[width=0.49\textwidth]{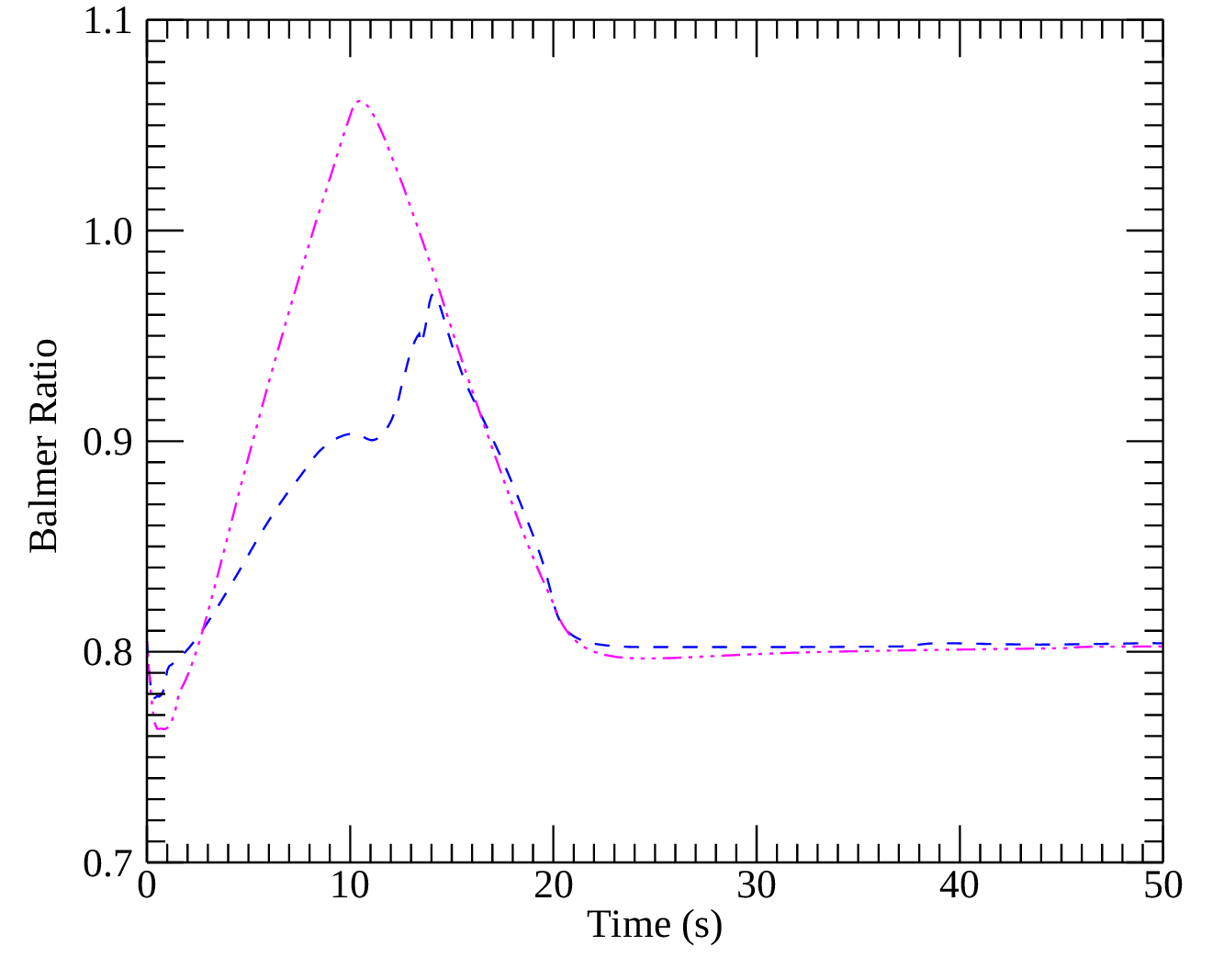}}
    \caption{$I_{c,rel}$ (top) and $R_B$ (bottom) as a function of time for the two cases $\mathrm{d4\_1.0e12\_20}$ (``case~1,'' dashed blue line) and $\mathrm{d3\_1.0e12\_25}$ (``case~2,'' dashed-triple-dotted magenta line).}
    \label{fig:events_c6684_bratio}
\end{figure}
\begin{figure}
    \includegraphics[width=0.49\textwidth]{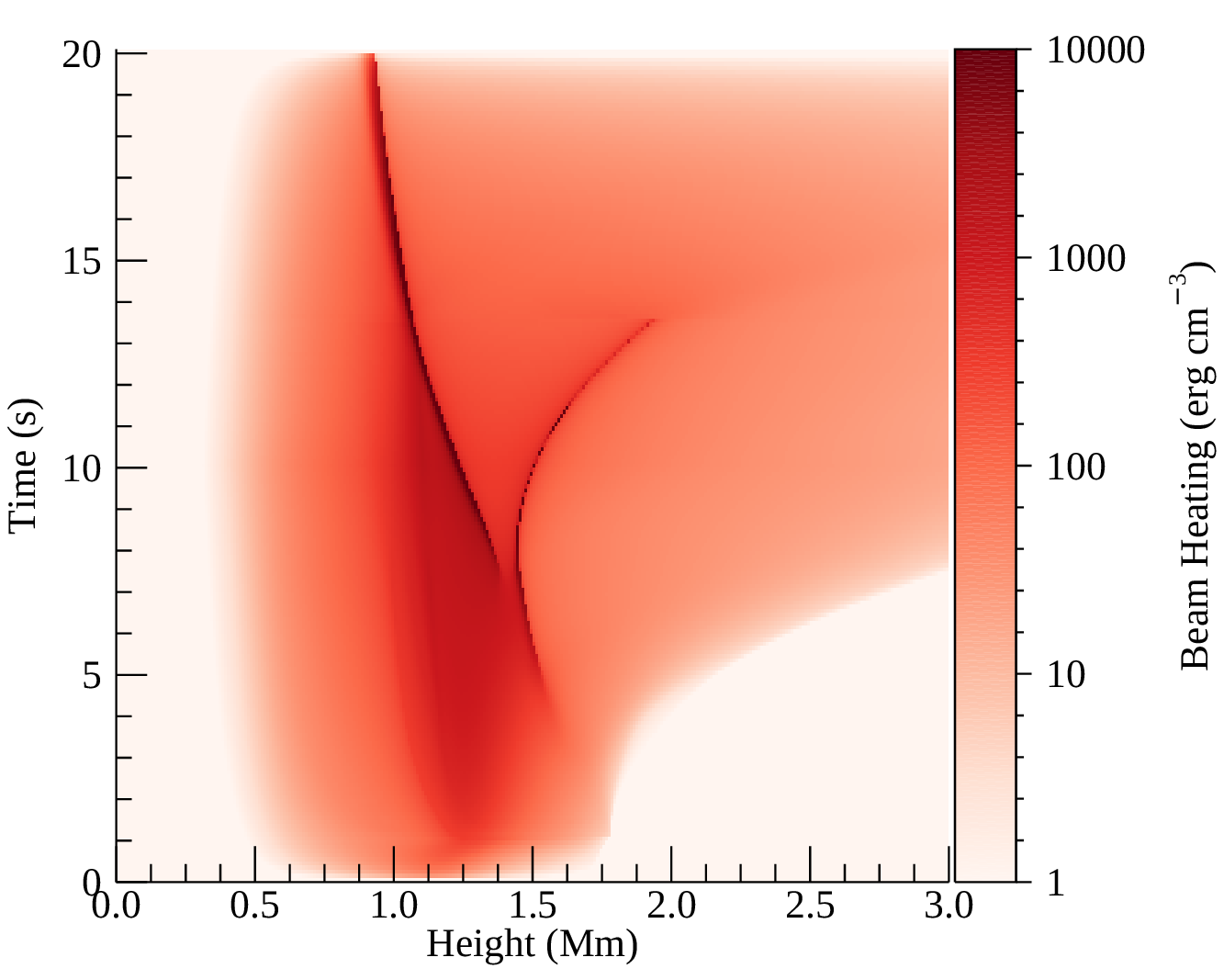}\\
    \includegraphics[width=0.49\textwidth]{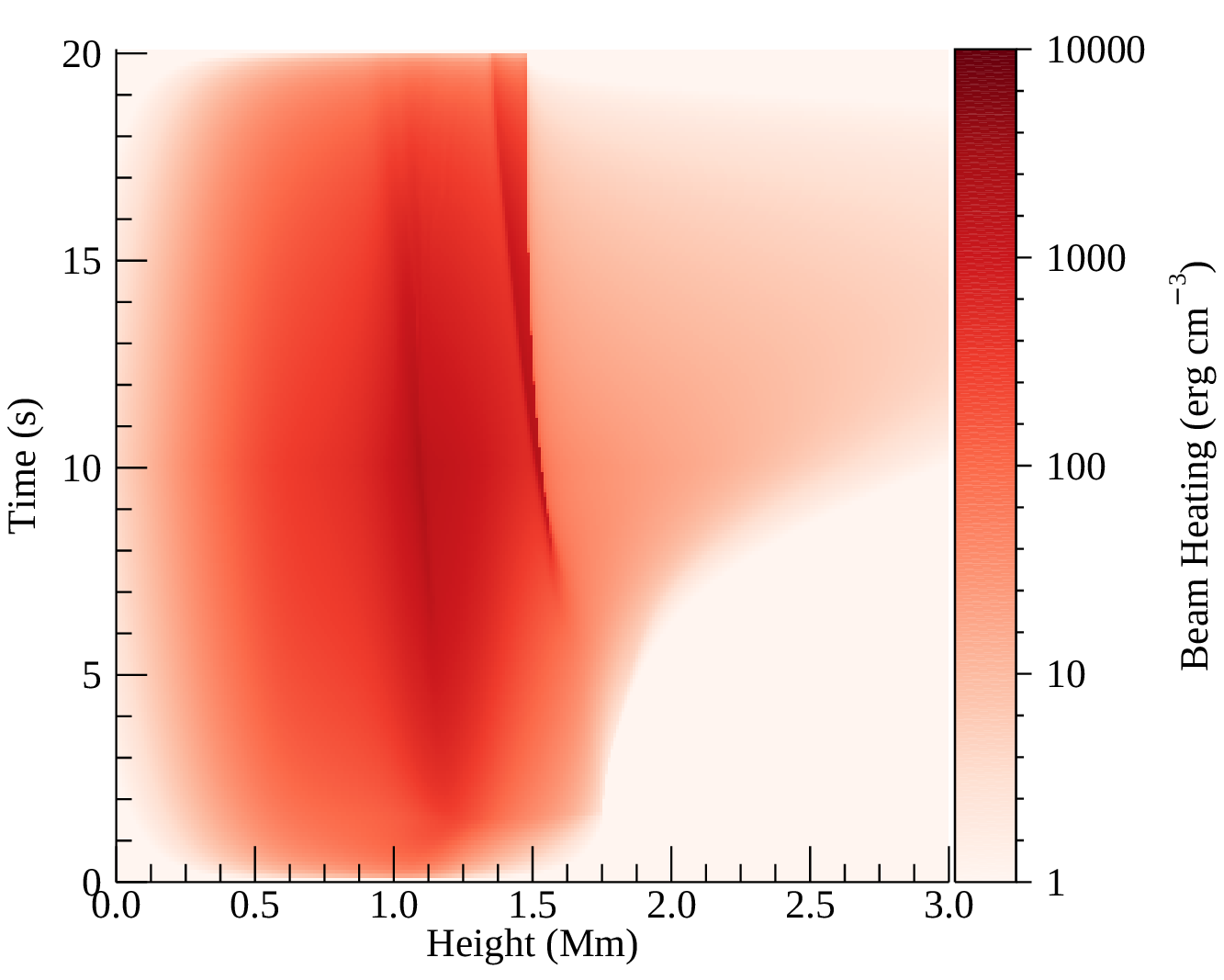}
    \caption{Beam heating term in the internal energy equation for case~1 (top panel) and case~2 (bottom panel) as a function of height and time. The scale is displayed as a color bar on the right side of each panel. The bin size in the space domain is 10~km, and 0.1~s in the time domain.}
    \label{fig:beam_heating}
\end{figure}
The first case we discuss in detail in this work has $I_{c,max} = 3.6$\% (see Fig.~\ref{fig:events_c6684_bratio}, top panel, dashed blue line). It has a spectral index of \mbox{$d = 4$}, a total beam energy of \mbox{$E_{tot} = 10^{12}$~erg~cm$^{-2}$}, and a lower-energy cutoff of \mbox{$E_c = 20$}~keV. It only shows a very short drop in continuum intensity ($<$~3~s). It reaches $I_{c,max}$ after \mbox{$t = 13.9$}~s, and decreases to an approximately constant level (which is still slightly above the pre-flare level) within \mbox{$t \approx 37$}~s after the start of the simulation.\\
\indent For this case, $R_{B,max} = 0.97$ (Fig.~\ref{fig:events_c6684_bratio}, bottom panel, dashed blue line), while $R_{P,max} = 1.01$. Similar to $I_{c,rel}$, $R_B$ shows only a short dip of~$<$~2~s. The increase happens in two stages, reaching the first (local) maximum around \mbox{$t = 10$}~s, and the second (global) maximum after \mbox{$t = 14.1$}~s, almost coincidental with the time of $I_{c,max}$. $R_B$ decreases rapidly thereafter and reaches pre-flare values after \mbox{$t \approx 21$}~s, i.e., shortly after the beam heating has ceased.\\
\begin{figure}
    {\includegraphics[width=0.49\textwidth]{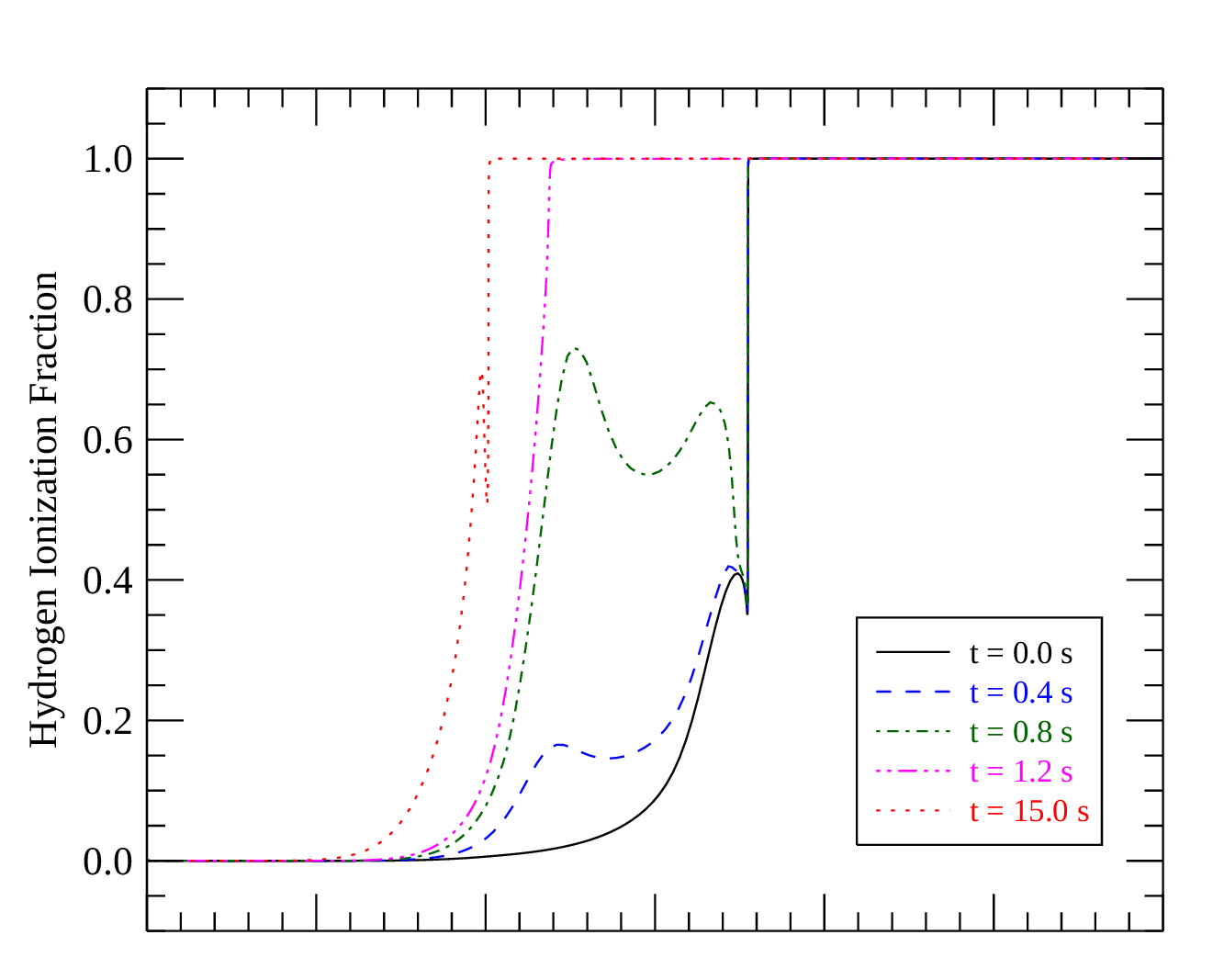}}
    {\includegraphics[width=0.49\textwidth]{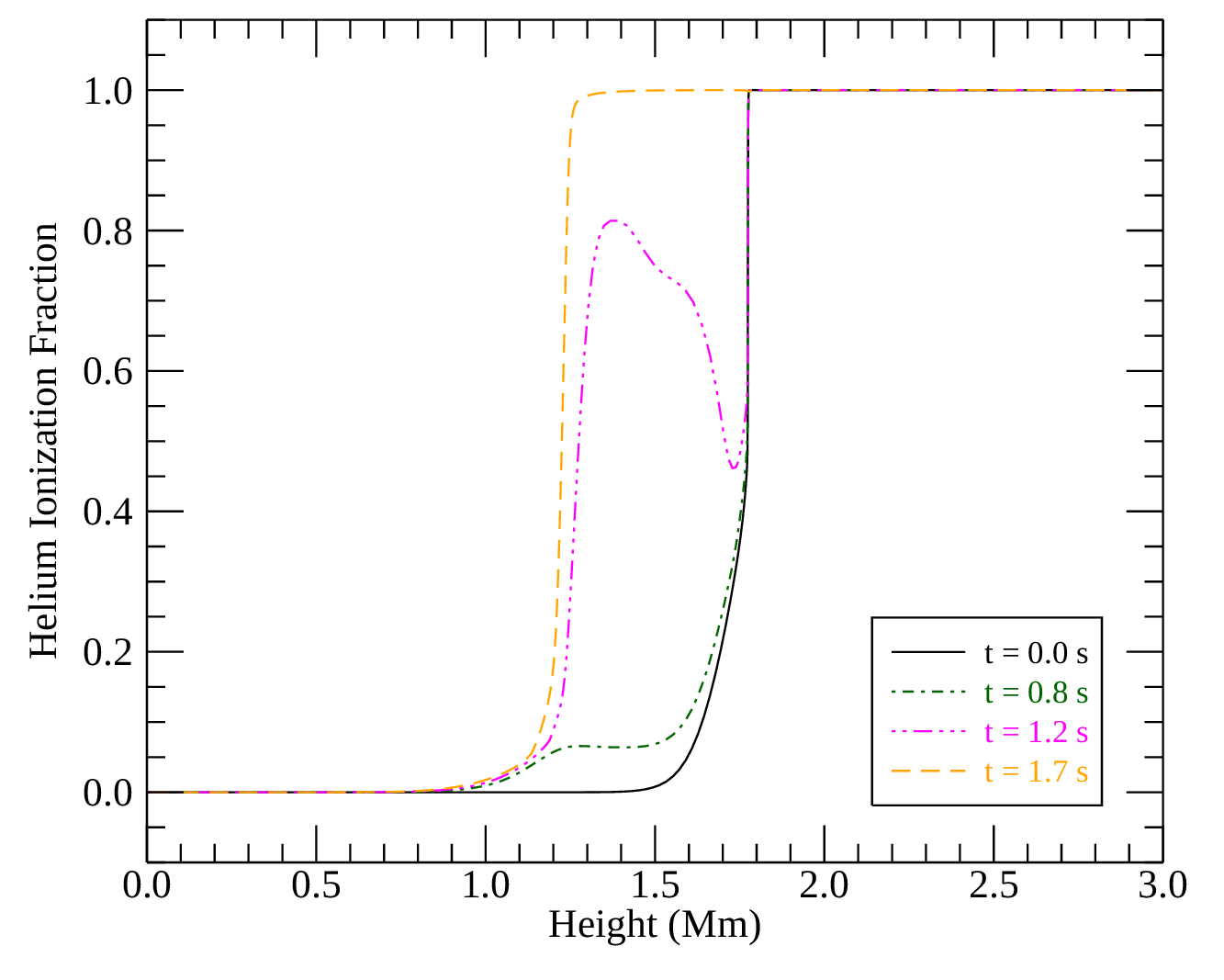}}
    \caption{Ionization fraction of hydrogen (top) and helium (bottom) as a function of height above the \mbox{$\tau_{500 \, nm} = 1$} photosphere at various times (indicated in the bottom right corner of each panel) for case~1.}
    \label{fig:ion_1}
\end{figure}
\begin{figure*}
    \resizebox{0.98\hsize}{!}
    {\includegraphics[width=0.99\textwidth]{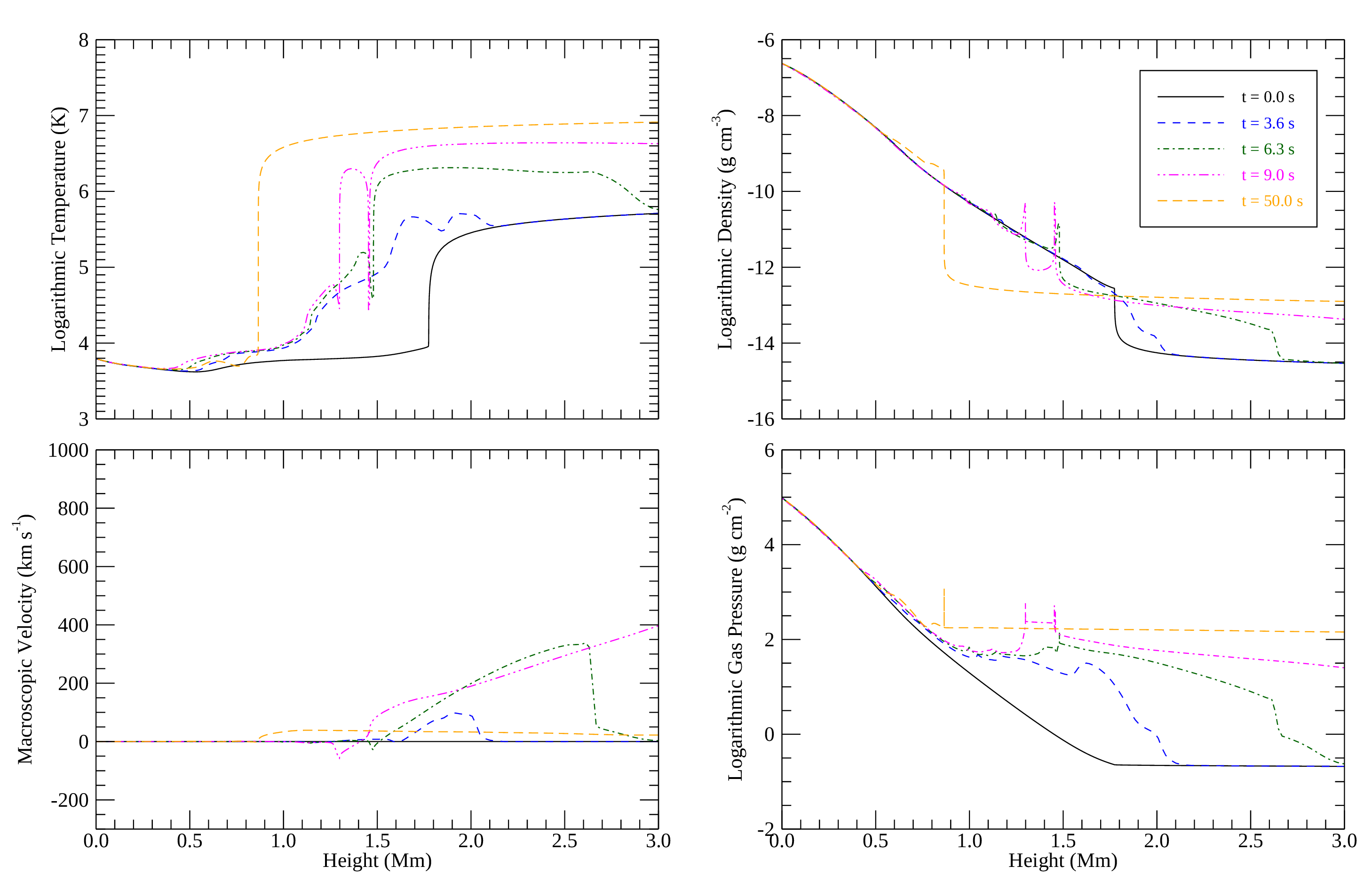}}
    \caption{Logarithmic temperature (top left), logarithmic density (top right), macroscopic velocity (bottom left) and logarithmic gas pressure (bottom right) as a function of height above the \mbox{$\tau_{500 \, nm} = 1$} photosphere at various times (indicated in the top right corner of the top right panel) for case~1.}
    \label{fig:combined_1}
\end{figure*}

\subsubsection{General description of case~1}\label{subsubsec:event_1_desc}

Most of the beam energy in the early stages of the flare is being deposited between \mbox{$h \approx 1.0$}~Mm and \mbox{$h \approx 1.75$}~Mm (middle to upper chromosphere; see Fig.~\ref{fig:beam_heating}, top panel). This beam heating heats the plasma in that region to between \mbox{$T = 10,000$~K} (near the upper and lower edges of these layers) and \mbox{$T = 17,000$}~K. At the same time, it causes an increasing number of hydrogen atoms to become ionized (see Fig.~\ref{fig:ion_1}, top panel), leading to a delicate balance of energy input and radiation that maintains the temperature constant for a few timesteps. Once most of the hydrogen has been ionized (at \mbox{$t = 1.1$}~s), it cannot effectively radiate away the energy input anymore, and the layers around \mbox{$h = 1.3$}~Mm see an increase in heating rate again. Helium now provides the main contribution to radiating away the energy input from the beam. As the bulk of the available helium gets ionized (Fig.~\ref{fig:ion_1}, bottom panel), the temperature stays approximately constant again, with the upper layers (up to \mbox{$h \approx 1.75$}~Mm) showing temperatures up to \mbox{$T \approx 10^5$}~K. Meanwhile, hydrogen ionization continues in an extended layer in the chromosphere (\mbox{$h \approx 0.9$}~Mm to \mbox{$h \approx 1.2$}~Mm, falling off exponentially toward the lower boundary) for multiple timesteps. The hydrogen ionization fraction decreases (i.e., the number of recombinations exceeds the number of ionizations) after the beam heating has reached its maximum and enters the decreasing phase. As can be seen in Fig.~\ref{fig:ion_1} (top panel, red dotted line), there is a small region within the transition region (now situated much further down in the atmosphere) where the ionization fraction drops rapidly. Late in the simulation, another short-lived episode of hydrogen ionization appears due to plasma that is moving down from the corona, hitting the layers around \mbox{$h \approx 1.0$}~Mm.\\
\indent Chromospheric evaporation (upward movement of chromospheric plasma along the magnetic loop) starts at \mbox{$t \approx 1.1$}~s (around the time when most of the helium has been ionized) and at an initial height of \mbox{$h \approx 1.8$}~Mm. This evaporation does not create enough overpressure \citep[see][]{1985ApJ...289..414F} to result in chromospheric condensation, since oxygen, carbon, and neon are still able to approximately balance the energy input from the beam via radiative cooling \citep{2018ApJ...852...61K}. However, at \mbox{$t \approx 3.0$}~s, a thermal runaway sets in, and the plasma gets heated to \mbox{$T \approx 10^6$}~K within 1~s. The resulting drastic increase in pressure is now enough to induce explosive evaporation and drive a chromospheric condensation downward. The downward movement sets in at \mbox{$t \approx 3.2$}~s, and about 200~km below the starting point of the evaporation (\mbox{$h \approx 1.6$}~Mm).\\
\indent Evaporation and condensation are visible as distinct peaks in the temperature profile (not as much in the density profile until \mbox{$t \approx 3.5$}~s), as seen in Fig.~\ref{fig:combined_1} (top left panel, dashed blue line), separated initially by a lower-temperature region which gets heated to similar temperatures over time. The evaporation shows an increased density by approximately one order of magnitude compared to the coronal plasma (Fig.~\ref{fig:combined_1}, top right panel, dashed-dotted green line). The upward velocity of the evaporating plasma increases toward the top of the simulation box (best seen in Fig.~\ref{fig:combined_1}, bottom left panel, dashed-triple-dotted magenta line), reaching \mbox{$v \approx 800$}~km~s$^{-1}$. This is a result of the higher sound speed in these regions of higher temperature. Meanwhile, the condensation moves at lower velocities, reaching a maximum of \mbox{$v \approx 50$~km~s$^{-1}$} (Fig.~\ref{fig:combined_1}, bottom left panel, dashed-triple-dotted magenta line). As it moves downward in the atmosphere, it accrues mass, increasing in density (compare the dashed-dotted green and the dashed-triple-dotted magenta line in Fig.~\ref{fig:combined_1}, top right panel). This increased density leads to more effective cooling, and the temperature decreases drastically to \mbox{$T < 40,000$}~K at \mbox{$t = 6.3$}~s. As the beam heating further increases, the region below the condensation reaches a point where the cooling via radiation cannot compensate for the increase in beam heating. At the local minimum in the cooling function, located between 50 and 100~km below the condensation, a local maximum of beam heating is formed. As a response, this region begins to heat up (from \mbox{$T \approx 1.4 \cdot 10^5$}~K to \mbox{$T \approx 2 \cdot 10^6$}~K in 4.1~s; see Fig.~\ref{fig:combined_1}, top left panel, dashed-dotted green and dashed-triple-dotted magenta line) and expand (decreasing in density by about one order of magnitude within 3~s). This results in an increase in density of the condensation above it, as well as the creation of a new high-density, low-temperature condensation below the expanding plasma (at height \mbox{$h \approx 1.35$}~Mm, starting at \mbox{$t \approx 7.3$}~s). The expansion of the region causes the upper (primary) condensation to reverse its direction of movement (now moving upward), and the lower (secondary) condensation to move downward. Both the upward-moving blob (since it cannot be called condensation anymore due to its upward movement) and the downward-moving secondary condensation have an increased density by about one order of magnitude compared to their surroundings, and compress as they travel through the atmosphere. The cool blob starts to diffuse at \mbox{$t \approx 13.4$}~s, decreasing drastically in density (and increasing in temperature). It reaches a density (and temperature) comparable with its surroundings within 0.2~s. Meanwhile, the secondary condensation dissipates much more slowly, possibly owing to its lower velocity (\mbox{$v \approx 60$}~km~s$^{-1}$).\\
\indent After \mbox{$t \approx 37.4$}~s, plasma coming from the corona at speeds of \mbox{$v \approx 150$}~km~s$^{-1}$ (either in the form of evaporation from the other footpoint itself or a rebound wave resulting from evaporation of the plasma in this footpoint hitting the evaporated plasma from the other footpoint at the top of the loop) hits the chromosphere at \mbox{$h \approx 0.95$}~Mm. This causes a slight increase in density and another minor episode of evaporation (and hydrogen ionization, as discussed earlier). Some plasma also continues its downward movement at speeds around \mbox{$v \approx 10$}~km~s$^{-1}$.

\subsubsection{White-light enhancements in case~1}\label{subsubsec:event_1_wl}

\begin{figure}
    \includegraphics[width=0.49\textwidth]{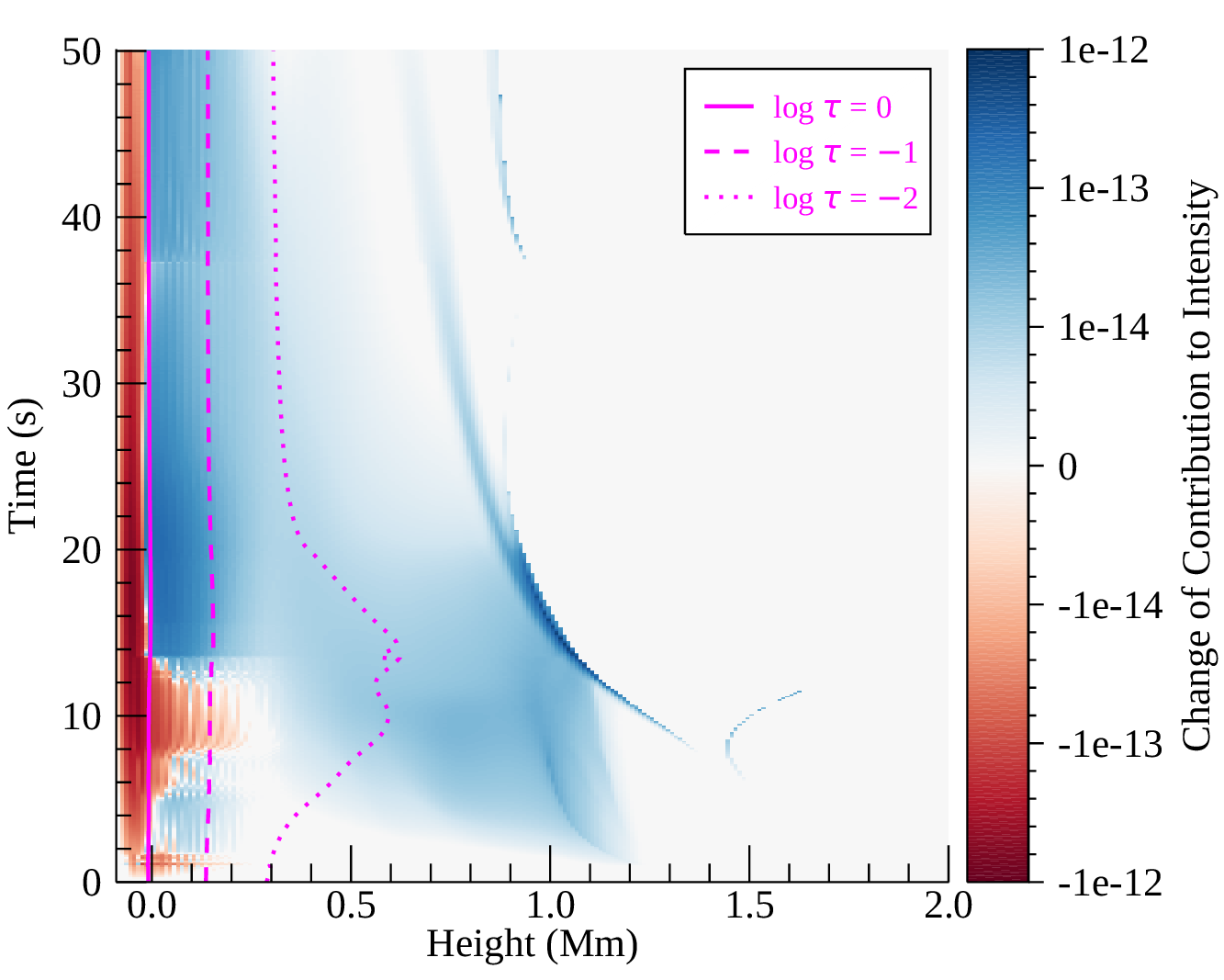}
    \caption{Change of the contribution to $I_c$ as a function of height and time on a logarithmic scale for case~1. The scale is displayed as a color bar on the right side. The bin size in the space domain is $10$~km, and $0.1$~s in the time domain. The solid, dashed, and dotted magenta lines indicate the height where \mbox{$\tau = 1.0$}, $0.1$, and $0.01$, respectively.}
    \label{fig:contrib_1}
\end{figure}
\begin{figure}
    \includegraphics[width=0.49\textwidth]{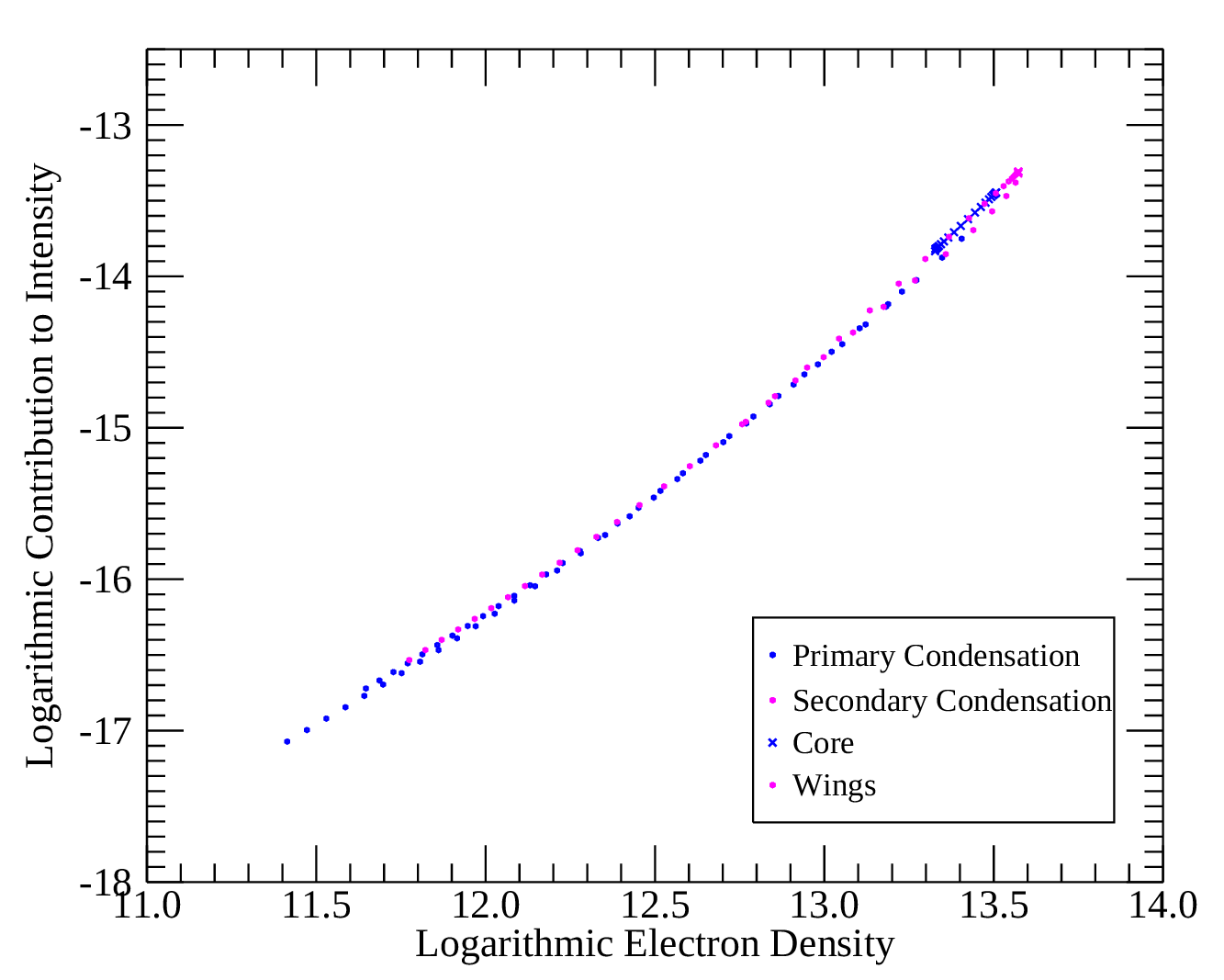}
    \caption{Logarithmic contribution to $I_c$ as a function of logarithmic electron density within the primary (blue) and secondary (magenta) condensation in case~1. Dots mark the wings, while crosses highlight the core of the condensations. The plot corresponds to \mbox{$t = 10.0$}~s.}
    \label{fig:contrib_edens_1}
\end{figure}
\indent In this simulation, the continuum contained 68.1\% of the total radiated energy. In Fig.~\ref{fig:contrib_1}, we show the change of the contribution function to $I_c$ (in logarithmic units) as a function of height (on the x-axis) and time (on the y-axis). Not visible in this plot is the fact that the main contribution overall comes from the photosphere. We start to see intensity enhancements in the chromosphere after \mbox{$t = 0.6$}~s, coinciding with the time when a significant portion of the hydrogen in this region has been ionized. The upper limit to these layers lies around \mbox{$h = 1.2$}~Mm, whereas the lower limit moves to deeper layers over time (following the hydrogen ionization). Overall, the largest contributions to the intensity excess come from the secondary condensation. A look at the (electron) densities in both condensation features shows that the contribution to $I_c$ seems to depend on the electron density (see Fig.~\ref{fig:contrib_edens_1}). This dependence is linear on a log-log scale throughout the condensations. Furthermore, the evolution of the contribution to $I_c$ as a function of electron density (available as movies upon request) suggests an electron density of roughly \mbox{$log(n_e) \gtrsim 13$} as a necessary, but not sufficient, condition for enhancements of $I_c$ to become significant. This explains why there is another episode of WL emission enhancement at a height of \mbox{$h \approx 0.9$}~Mm starting at \mbox{$t \approx 38$}~s, which is precisely at the time and location where the short-lived episode of hydrogen ionization due to the impact of material coming from the corona happens (see Sect.~\ref{subsubsec:event_1_desc}).\\
\indent The contribution to $I_c$ from layers below the photosphere decreases substantially during the simulation. This is due to the increase in the opacity, exemplified in Fig.~\ref{fig:contrib_1} with a solid magenta line marking the $\tau = 1$ layer. Although this layer does not vary much in height throughout the simulation, this small change is enough in order for a layer of a given opacity to be shifted to heights with a considerably lower temperature and therefore source function, which decreases the contribution to $I_c$.\\
\indent There is a noticeable decrease of the contribution function in photospheric layers (up to the TMR) between $t \approx 6$~s and $t \approx 14$~s. The cause of this can be understood by looking at the dotted magenta line in Fig.~\ref{fig:contrib_1}, which marks the height of the \mbox{$\tau = 0.01$} layer. It reveals an increase in opacity in chromospheric regions (and down to the TMR and below). This increase is mainly due to an elevated population in the $n = 2$ states of hydrogen. Following the description in Sect.~\ref{subsec:statistics}, portions of the photospheric radiation are absorbed in these layers of enhanced opacity (relative to the pre-flare level). We discuss this in more detail in Sect.~\ref{sec:discussion}. The magenta curves in Fig.~\ref{fig:contrib_1} furthermore show that the excess emissions in chromospheric layers stem from regions in the atmosphere that can be characterized as optically thin. As mentioned before, the \mbox{$\tau = 1.0$} layer, marking the change to an optically thick regime, does not change significantly in height throughout the simulation.\\
\begin{figure}
    \includegraphics[width=0.49\textwidth]{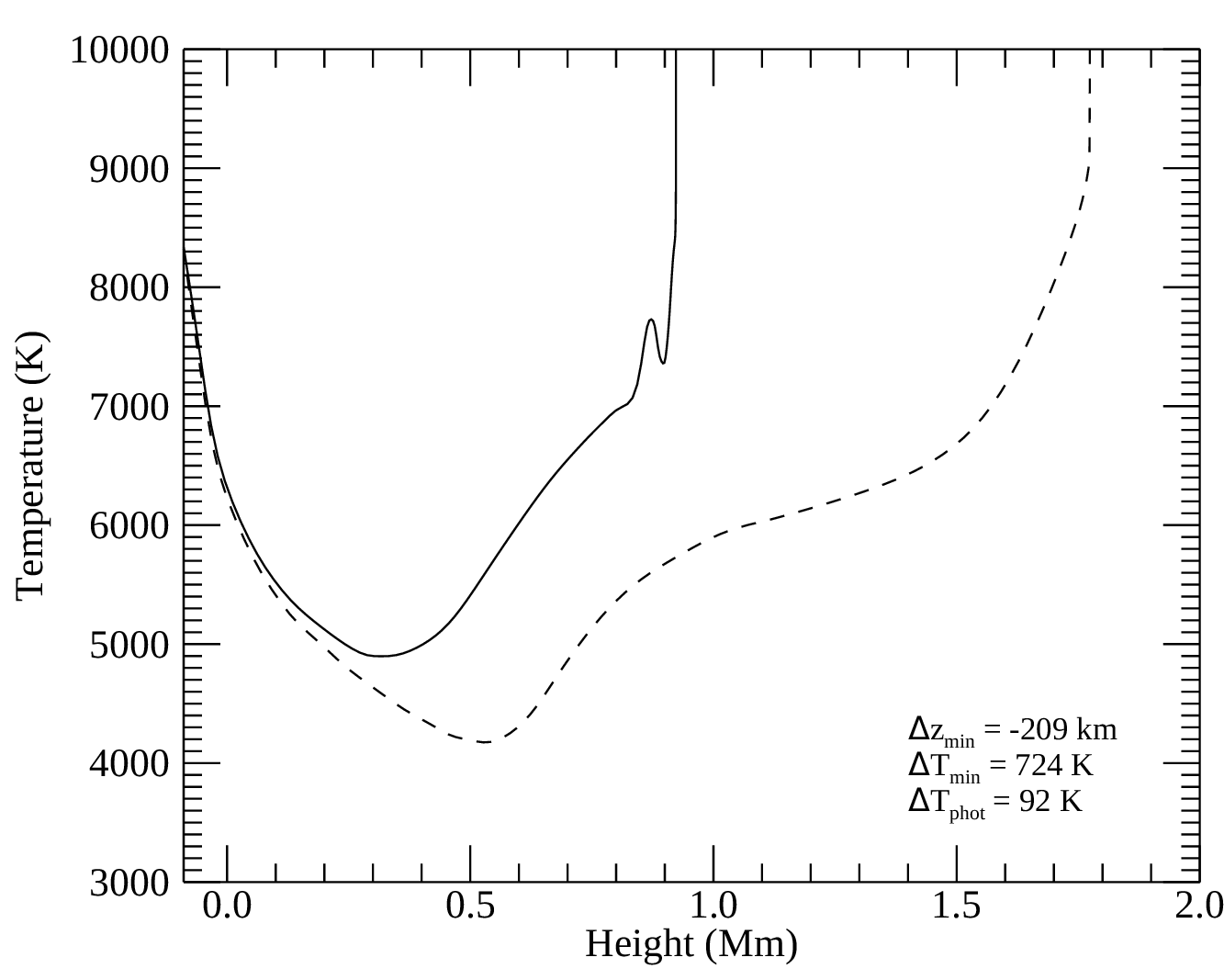}
    \caption{Temperature around the TMR at \mbox{$t = 20.2$}~s (solid line), i.e.,~the time of highest temperature in the TMR, and at \mbox{$t = 0$}~s (dashed line) for case~1. The bottom right indicates the height difference between the TMR at \mbox{$t = 0$}~s and at \mbox{$t = 20.2$}~s, as well as the temperature differences in the TMR and photosphere at these times.}
    \label{fig:TMR_1}
\end{figure}
\indent In this simulation, the TMR and even the photosphere (defined here as the layer where \mbox{$\tau_{500 \, nm} = 1$}) get heated. As Fig.~\ref{fig:TMR_1} indicates, at the time of maximum temperature difference (compared to \mbox{$t = 0$}~s), i.e.,~at \mbox{$t = 20.2$}~s, the TMR is 724~K hotter and lies~$>$~200~km deeper. The photosphere reaches its maximum temperature a little earlier (at \mbox{$t = 17.1$}~s), with the temperature enhancement being 94~K. At the end of the simulation, the TMR and photosphere are still 363~and~27~K hotter, respectively, than at \mbox{$t = 0$}. Note that beam heating at the height of the TMR can only take place through very high-energy electrons and is therefore very low. The main reason for the temperature increase is as follows: Some of the (predominantly) Balmer-continuum photons created in the chromosphere through recombination are traveling downward in the atmosphere and are absorbed in the regions around the TMR and below. This process is known as radiative backwarming. This decreases the overall radiative losses in the photosphere, resulting in heating of the TMR and photosphere. This heating is then also the reason for the enhanced photospheric emissions that become apparent in Fig.~\ref{fig:contrib_1} after $t \approx 14$~s. We note that the heating of the photosphere via backwarming commences already early in the simulation, once the majority of the hydrogen has been ionized and recombination rates are high. The change of the contribution to $I_c$ from the photosphere, however, remains negative until $t \approx 14$~s due to the aforementioned increase in the opacity in the chromosphere, and the change only becomes positive once the heating is high enough, and subsequently when the opacity in the chromosphere decreases again. In fact, as the contribution from hydrogen recombination in the chromosphere declines, the percentage of photospheric contribution to the total change of the contribution function (as depicted in Fig.~\ref{fig:contrib_1}) increases from $<$10\% at $t \approx 13$~s to $>$80\% at $t \approx 20$~s.


\subsection{Case~2: High Balmer ratio}\label{sec:event_2}

\indent Case~2 is the case with the highest $R_{B,max}$ (with $R_{B,max} = 1.06$, see Fig.~\ref{fig:events_c6684_bratio}, bottom panel, dashed-triple-dotted magenta line, while $R_{P,max} = 1.02$). It has a spectral index of \mbox{$d = 3$}, a total beam energy of \mbox{$E_{tot} = 10^{12}$~erg~cm$^{-2}$}, and a lower-energy cutoff of \mbox{$E_c = 25$}~keV. $R_{B,max}$ in this simulation is reached after \mbox{$t = 10.5$}~s, which is 3.6~s earlier than in case~1. The decrease after the maximum is again roughly linear and very similar in shape to case~1.\\
\indent As shown in Fig.~\ref{fig:events_c6684_bratio} (top panel, dashed-triple-dotted magenta line), $I_{c,max} = 4.0$\% for this case. It is reached after \mbox{$t = 14.2$}~s, almost coincident with case~1. Generally, the light curve of $I_{c,rel}$ is smoother than it is for case~1. The most obvious feature, besides the main maximum, is a deep and long-lasting decrease in intensity from the start of the simulation until \mbox{$t = 7.4$}~s, with the peak decrease being over 2.1\% at \mbox{$t = 3.5$}~s.
\begin{figure}
    {\includegraphics[width=0.49\textwidth]{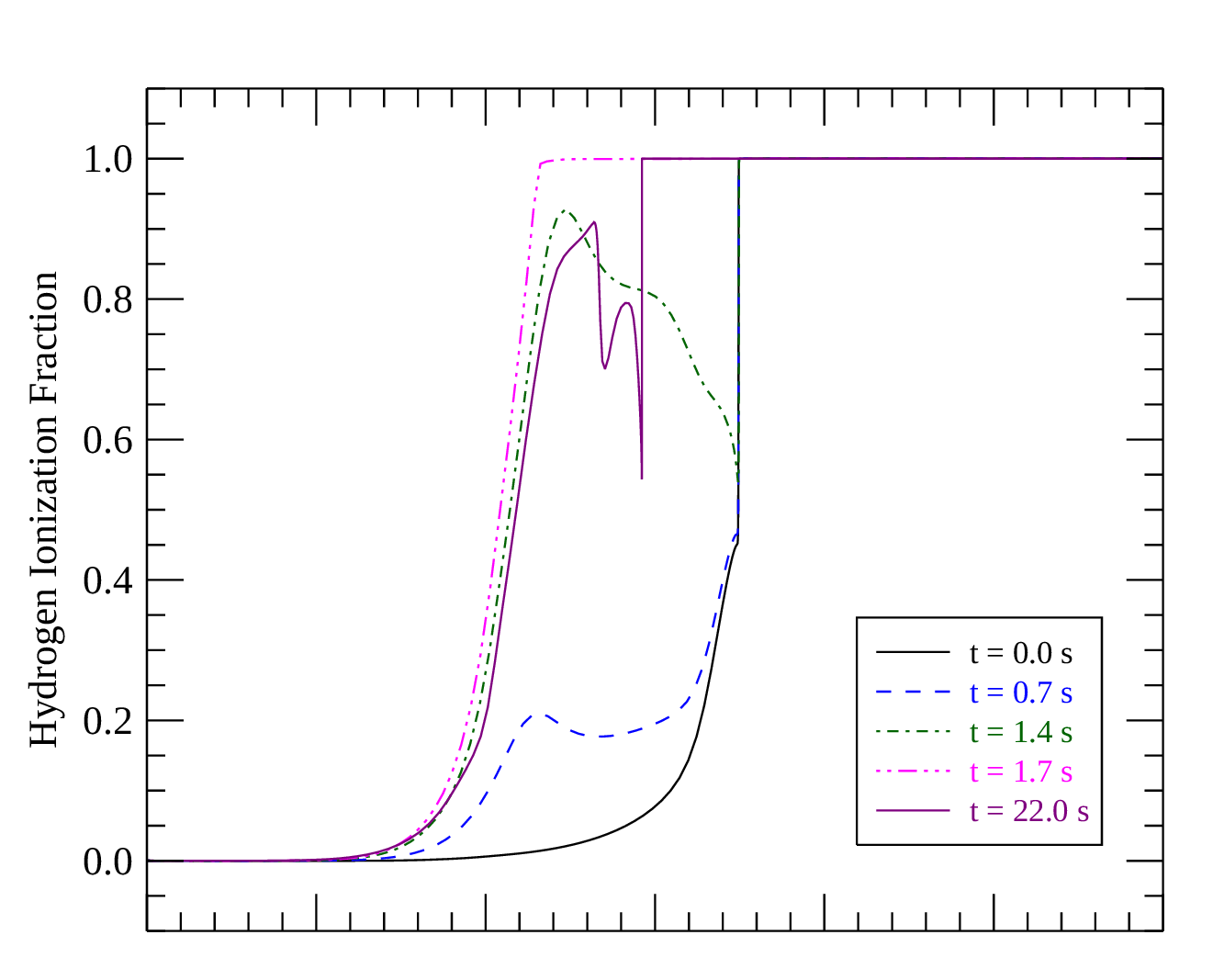}}
    {\includegraphics[width=0.49\textwidth]{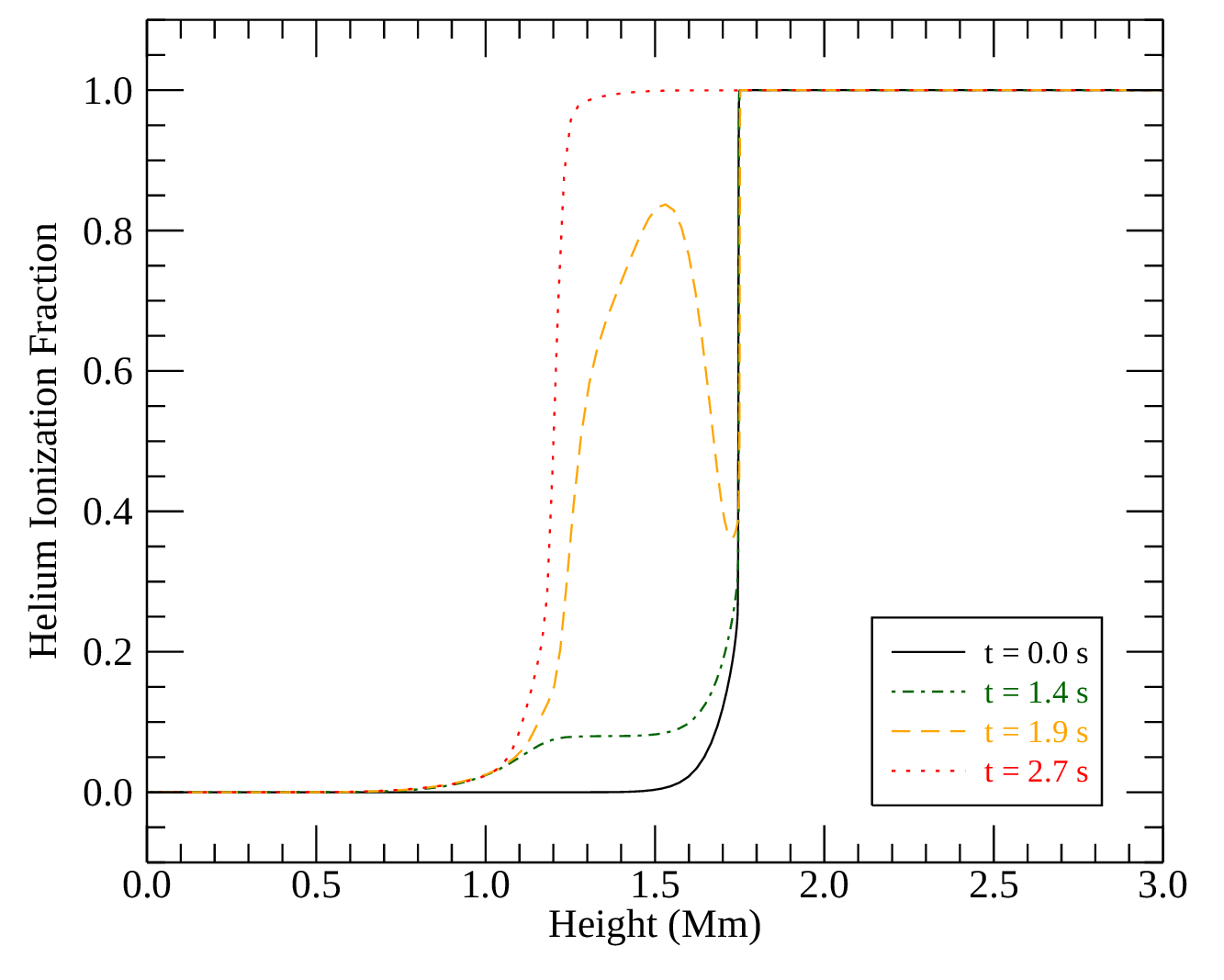}}
    \caption{Same as Fig.~\ref{fig:ion_1}, but for case~2.}
    \label{fig:ion_2}
\end{figure}
\begin{figure*}
    \resizebox{0.98\hsize}{!}
    {\includegraphics[width=0.99\textwidth]{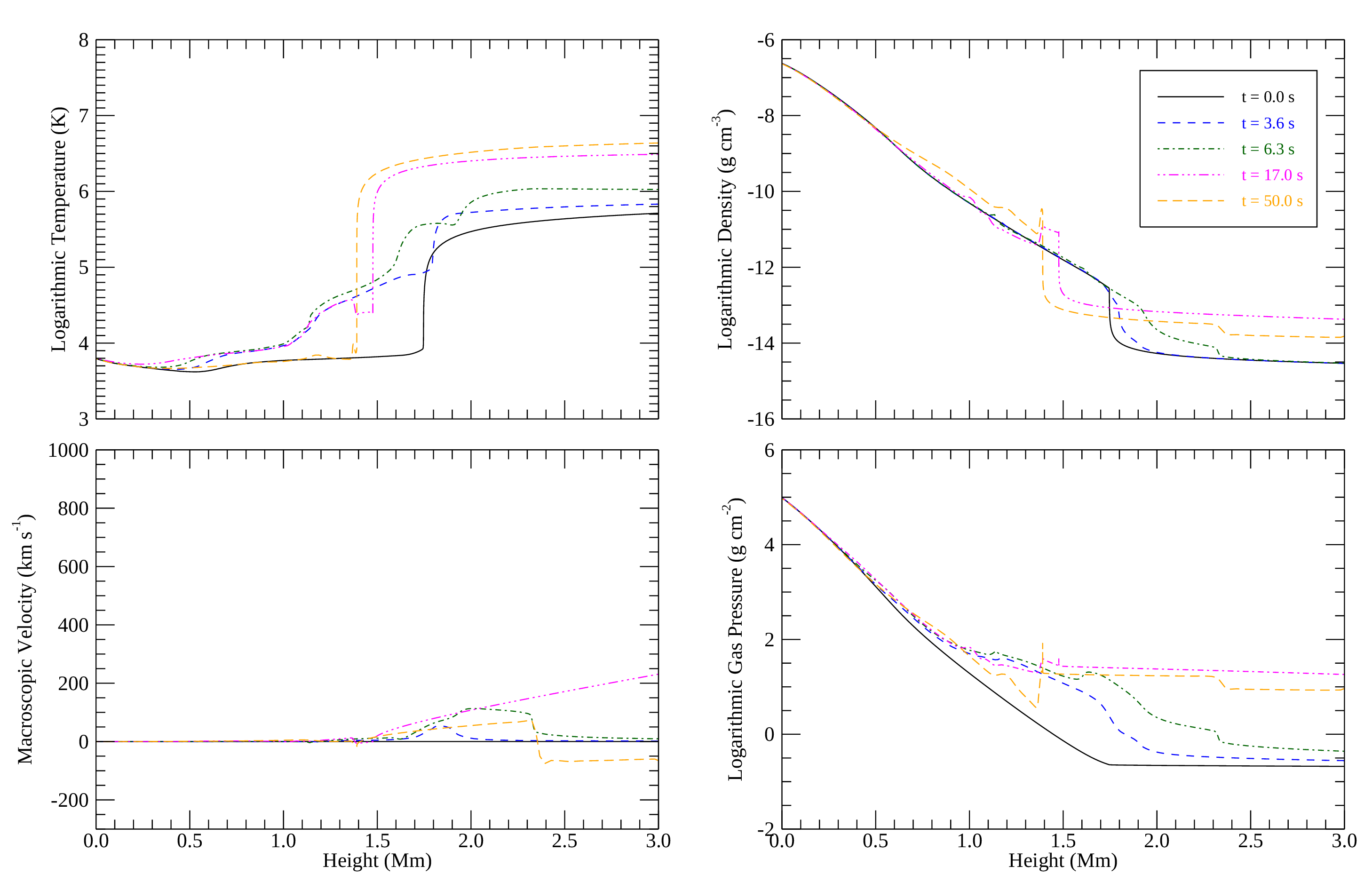}}
    \caption{Same as Fig.~\ref{fig:combined_1}, but for case~2.}
    \label{fig:combined_2}
\end{figure*}

\subsubsection{General description of case~2}\label{subsubsec:event_2_desc}

The location of beam energy deposition is initially similar to case~1 (see Fig.~\ref{fig:beam_heating}, lower panel), but the beam penetration depth is generally deeper. Within the first 1.3~s (after which most of the hydrogen has been ionized, see Fig.~\ref{fig:ion_2}, top panel, dashed-dotted green line), the plasma between \mbox{$h \approx 1.0$}~Mm and \mbox{$h \approx 1.75$}~Mm gets heated to between \mbox{$T = 11,000$}~K (near the lower edges) and \mbox{$T = 15,000$}~K. After that, the layers between \mbox{$h = 1.2$}~Mm and \mbox{$h = 1.7$}~Mm get heated further and faster, to \mbox{$T \approx 25,000$}~K within 0.3~s. Thereafter, helium ionization balances the energy input. Simultaneously, we see continuing hydrogen ionization deeper down between \mbox{$h = 0.8$}~Mm and \mbox{$h = 1.15$}~Mm. Similarly to case~1, the hydrogen ionization fraction decreases once the beam heating starts to decrease. However, as shown in Fig.~\ref{fig:ion_2} (top panel, solid purple line), the region where the ionization fraction rapidly drops is much more extended (in height) than in case~1, where it showed an extent of~$<$~100~km. In case~2, this region is almost 500~km thick. Furthermore, the density in this region is \mbox{$1-2$}~orders of magnitude lower than in case~1. Late in the simulation, similar to case~1, there is a short spike in hydrogen ionization at \mbox{$t \approx 45.4$}~s, when plasma coming down from the corona hits the chromospheric regions.\\
\indent Chromospheric evaporation in this simulation starts later than in case~1, around \mbox{$t \approx 1.8$}~s (when all of the hydrogen and most of the helium in the upper chromosphere have been ionized). The location, however, is similar (\mbox{$h \approx 1.8$}~Mm). Furthermore, there are again two distinct phases of evaporation, an initial, gentle one resulting in a lower plasma velocity (\mbox{$v \approx 70$}~km~s$^{-1}$ when first crossing \mbox{$h = 2.0$}~Mm) and temperature (\mbox{$T \approx 10^5$}~K), and another delayed, explosive one, producing plasma velocities comparable to case~1 (\mbox{$v \approx 100$}~km~s$^{-1}$ when first crossing \mbox{$h = 2.0$}~Mm) and temperatures reaching \mbox{$T > 10^6$}~K. The later evaporation catches up with the earlier one at a height of \mbox{$h \approx 3.5$}~Mm. The maximum velocity of the evaporation, however, is about $200$~km~s$^{-1}$ lower than in case~1 (\mbox{$v \approx 600$}~km~s$^{-1}$).\\
\indent Chromospheric condensation takes about twice as long to manifest itself compared to the previous simulation (\mbox{$t \approx 6.2$}~s) due to the longer time it takes until the second, explosive phase of evaporation happens. The initial height of the condensation is similar to case~1, as is the temperature. However, the condensation only reaches a velocity of \mbox{$v \approx 30$}~km~s$^{-1}$. Contrary to case~1, the density and temperature profiles of the condensation broaden over time and develop spiky edges (with even lower temperature and higher density; see Fig.~\ref{fig:combined_2}, top left and top right panel, dashed-triple-dotted magenta line). However, the edges disappear over time, and the whole profile flattens out as the condensation slowly dissipates (without the creation of an expanding, low-density region below it, and without a secondary condensation).\\
\indent Since the evaporation is slower, it also takes longer until the evaporation from the other footpoint reaches this footpoint (or, in a different interpretation, until the rebound wave hits these regions). This happens at \mbox{$t \approx 45.4$}~s, compared to \mbox{$t \approx 37.4$}~s for case~1. The site of this impact is much higher up in this simulation (\mbox{$h \approx 1.5$}~Mm, compared to \mbox{$h < 1.0$}~Mm for case~1), at a location with lower (electron) density. Nevertheless, we do see the same rebound effect and the same creation of a cool blob moving downward as a result of the impact.

\subsubsection{White-light enhancements in case~2}\label{subsubsec:event_2_wl}

\begin{figure}
    \includegraphics[width=0.49\textwidth]{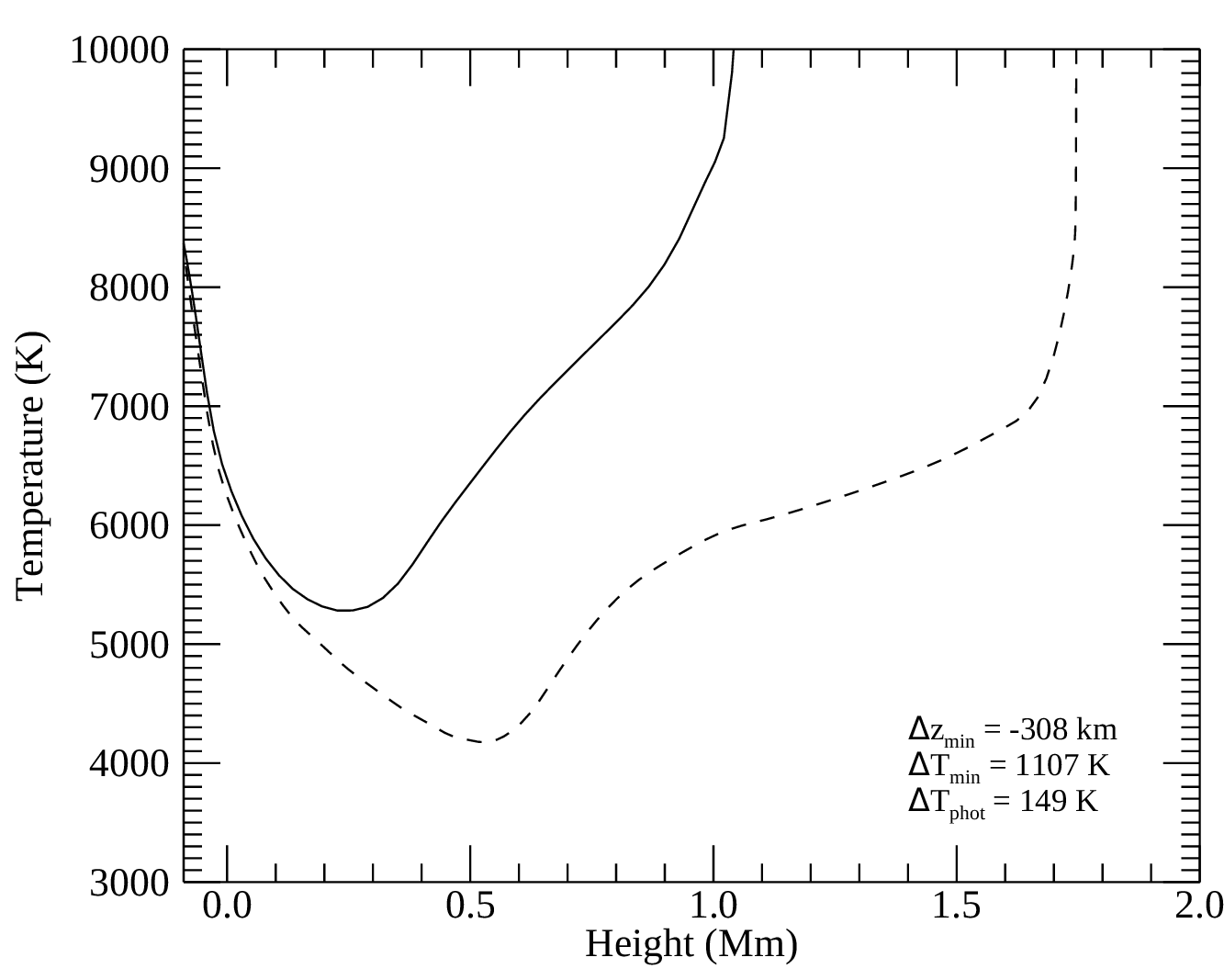}
    \caption{Same as Fig.~\ref{fig:TMR_1}, but for case~2 at \mbox{$t = 17.0$}~s (the time of highest temperature in the TMR in this simulation).}
    \label{fig:TMR_2}
\end{figure}

The continuum harbored 54.3\% of the total radiated energy, significantly less than for case~1. The most apparent difference in the $I_{c,rel}$ light curve between case~1 and case~2 is the significantly more pronounced initial decrease in intensity for the latter. Since the duration between the start of the simulation and the re-brightening of the continuum is determined by the ratio of recombinations to photoionizations \citep{2006ApJ...644..484A}, this means that the number of free electrons available for recombinations reaches levels high enough for the continuum to re-brighten earlier in case~1 than case~2. This suggests weak heating rates (see Sect.~\ref{subsec:statistics}) in this simulation compared to case~1, which we comment on in Sect.~\ref{sec:discussion}.\\
\begin{figure}
    \includegraphics[width=0.49\textwidth]{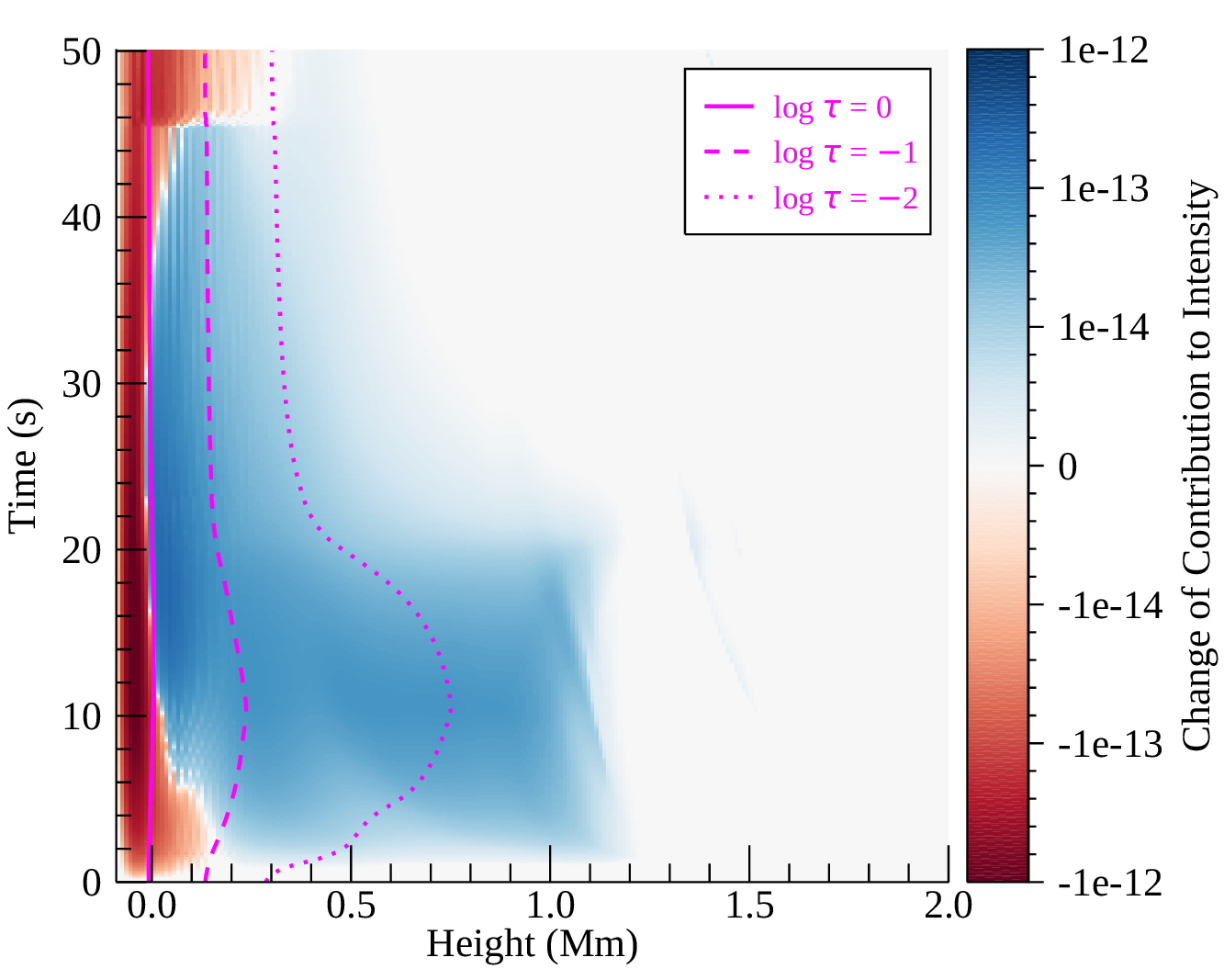}
    \caption{Same as Fig.~\ref{fig:contrib_1}, but for case~2.}
    \label{fig:contrib_2}
\end{figure}
The time until there appear contributions from somewhere other than the photosphere is comparable to case~1. These contributions (between \mbox{$h \approx 0.2$ Mm} and \mbox{$h \approx 1.2$ Mm}) are more extended than in case~1, more uniform, and have a higher contribution overall (except for the condensations discussed in case~1). Chromospheric condensation does not produce significantly higher contributions to the continuum intensity, owing to the lower electron density compared to the condensations in case~1. The contribution from below the photosphere is decreased throughout the whole simulation time, as in case~1 (again due to the higher opacity). The increase in opacity, however, is larger than in case~1, which is the reason why the continuum light curve shows a more pronounced decrease. Similar to case~1, the height of the \mbox{$\tau = 1$} layer (Fig.~\ref{fig:contrib_2}, solid magenta line) does not change much throughout the simulation, and the overwhelming part of the excess WL emissions can therefore comfortably be characterized as optically thin.\\
\indent As in case~1, both the TMR and the photosphere experience heating. As displayed in Fig.~\ref{fig:TMR_2}, the maximum temperature enhancement of the TMR is 1107~K (at \mbox{$t = 17.0$}~s, so about 4~s earlier than in case~1) and it lies~$>$~300~km deeper, exceeding the respective values in case~1. The photosphere also reaches its maximum temperature earlier (at \mbox{$t = 16.3$}~s), with the temperature enhancement being 150~K. At the end of the simulation, the TMR is still 458~K hotter than at \mbox{$t = 0$}, whereas the photosphere is 18~K cooler. Beam heating at the TMR height again is low. Backwarming once more serves as the main heating mechanism in these layers. The higher temperatures in case~2 follow from a beam energy deposition that is more spread-out (see Fig.~\ref{fig:beam_heating}; for a discussion, see below) than in case~1. This results in a larger area where hydrogen recombination can happen, and the resulting chromospheric contribution to the total WL emission enhancement is bigger. More photons can participate in backwarming, which subsequently increases heating in the photosphere and, therefore, the resulting temperatures.\\
\indent Similar to case~1, the photospheric contribution to the total change of the contribution function increases over time. At $t = 10$~s, the photosphere contributes about 22\% to the total enhancements. At the time of $I_{c,max}$, roughly 46\% of the intensity enhancement stems from the photosphere. During the declining phase of $I_{c,rel}$, this share rises to $>$80\%.


\section{Discussion}\label{sec:discussion}


Electron beams with a higher low-energy cutoff and a lower spectral index contain more high-energy electrons that can penetrate deeper into the solar atmosphere. Assuming that the excess WL emissions come from deeper layers, it would then be straightforward to presume that electron beams with these specifications would be more likely to create an atmospheric response favorable for the creation of WL enhancements. The results presented in Fig.~\ref{fig:fchroma_params} agree with this notion to some degree, as the simulations with the lowest cutoff energy and high spectral indices do not show WL enhancements. We note, however, that the highest ($=25$~keV) cutoff energies do not lead to an increase in excess WL probability. Comparing the different beam parameters and how the beam heating is distributed across the domain, we conclude the following: (i) A higher total beam energy naturally results in more energy being deposited throughout the atmosphere, (ii) a higher cutoff energy results in a deeper penetration of the beam electrons, and (iii) a higher spectral index results in a beam penetrating less deeply. We note that there is a particularly significant jump from spectral index $d = 3$ to $d = 4$. These conclusions, together with Fig.~\ref{fig:fchroma_params}, are in line with the picture where the WL enhancements do not necessarily stem from the deepest regions in the solar atmosphere, but rather from chromospheric layers. However, many of the simulations with high total beam energies and high spectral indices and/or low cutoff energies are not (yet) included in the F-CHROMA grid. This is due to the fact that the timesteps within these simulations reach values on the order of $10^{-6}$~s, making them very time-consuming and computationally expensive. Nonetheless, more finished simulations with a total beam energy of $10^{12}$~erg~cm$^{-2}$ are essential to confirm the above conclusions.\\
\indent The continuum light curve at 6684~\AA~shows a deeper and more extended (in time) dip at the beginning of the simulation for case~2 compared to case~1. This is counterintuitive, since case~2 has a lower spectral index and higher cutoff energy, and should therefore, following the reasoning above, provide a higher flux of high-energy electrons reaching the lower layers. However, following Fig.~\ref{fig:beam_heating}, we can understand this in the following way: As discussed above, a lower spectral index together with a higher cutoff energy results in a beam that penetrates deeper into the atmosphere. Therefore, the beam in case~2 raises the $n = 2,3,...$ populations of hydrogen down to lower layers than in case~1. The result is a stronger increase in opacity deeper down (at heights of the pre-flare TMR and below) in case~2, combined with an opacity increase at chromospheric layers that is comparable to that in case~1. At the same time, the deeper penetration leaves less energy to be deposited at the higher layers, which therefore ionize more slowly. This means that (i) they remain opaque to photospheric radiation for longer, and (ii) the increase in contributions from chromospheric layers within the first approximately 7~seconds is not high enough to outweigh the decrease in contributions from the photosphere and below, resulting in a decrease in overall intensity. In a similar interpretation, the increased opacity increases the formation height of layers with a given $\tau$. At these higher altitudes, the temperature is lower, resulting in a lower source function and, consequently, an intensity decrease. In case~1, the (sub-) photospheric emission decreases seen early in the simulation are not as pronounced, and the overall dimming is shallower and shorter.\\
\indent Dimmings in the continuum light curve as portrayed in this paper have so far been elusive in observations. We speculate that the reason could be the viewing angle. In simulations like those produced by the RADYN code, the magnetic loop is oriented along the line of sight. As discussed, there is an increase in opacity above the photosphere within the loop, which absorbs photons coming from below and therefore decreases the photospheric contribution to the intensity. At the same time, if the emission enhancements from optically thin hydrogen recombination early on are not big enough to compensate for these absorbed photons, the result is an overall dimming in the continuum light curve. In observations, however, flaring loops are not oriented exclusively along the line of sight. Additional projection effects may then result in (at least parts of) the photospheric emission bypassing the cloud of higher opacity above it in the loop, allowing the photospheric contribution to still be observed. The increase in chromospheric contributions then leads to an increase in continuum intensity without a prior dimming.\\
\indent The strong correlation between the Balmer and Paschen ratios suggests that they are created by the same mechanism. The relationship between the Balmer (and Paschen) ratio and the continuum intensity at 6684~\AA~can be explained as follows: The Paschen continuum is the main contributor to the optically thin continuum radiation at 6684~\AA. A higher Paschen ratio (i.e., a higher intensity of the continuum blueward of the Paschen jump relative to the redward continuum), therefore, naturally results in a higher continuum intensity at 6684~\AA. Consequently, since the Balmer and Paschen continua are created by the same mechanism (only differing in the final energy state in which the electron in the neutral hydrogen atom finds itself), a higher Balmer ratio also implies a higher continuum intensity at 6684~\AA.\\
\indent The contribution function to intensity for the continuum at 6684~\AA~shows that for both cases studied in detail in this work, the majority of the intensity increase during the maximum of the WL enhancement stems from optically thin layers. In case~1, the largest contribution to the continuum increase comes from the primary and secondary chromospheric condensations. The linear relationship between the logarithmic contribution function to intensity and the logarithmic electron density in these features supports the notion that optically thin H-recombination radiation is responsible for the continuum excess \citep{2017ApJ...836...12K}. This also explains the lower contribution to intensity that stems from the condensation in case~2, as it has a lower electron density than both condensations in case~1.\\
\indent The role of increased H$^-$ emission is similar in both cases, as the photosphere provides the major contribution to the increase in continuum intensity in the declining phase of the intensity enhancement (when the additional chromospheric contribution from hydrogen recombination starts to fade away). The higher temperatures around the TMR and in the photosphere in both case studies are due to backwarming, predominantly in the Balmer continuum. In this scenario, the elevated temperature increases the Planck function and, by extension, the source function. This increase causes excess H$^-$ emission.\\
\indent We stress the fact that even the simulations with the most energetic electron beams are not able to reproduce observationally measured white-light enhancements. Moreover, we have shown that on average, about half of the total radiated energy in WL cases is contained within the continuum, which is significantly below the calculations of \cite{2011A&A...530A..84K}, who found the continuum to make up two-thirds of the radiated energy in solar flares. In only six WL cases in the F-CHROMA grid did the continuum harbor more than 60\% of the total radiated energy. This leads us to conclude that other energy input mechanisms (like proton beams or Alfvén waves) are needed to explain observations.


\section{Summary and conclusion}


The F-CHROMA grid contains (as of the time of writing) 84~flare simulations, of which 45 show at least some enhancements in the 6684~\AA~continuum. Applying a threshold of 0.1\% for the relative continuum intensity increase (compared to the pre-flare level), 33 can be classified as white-light cases. 12 cases even exceed a relative intensity increase of 1\%.\\
\indent The total energy (or maximum beam flux) of the electron beam impacting the lower atmosphere is the most decisive factor in determining whether WL intensity enhancements will be present or not. Both a higher low-energy cutoff and a lower spectral index result in a beam spectrum containing a higher fraction of high-energy electrons, which can penetrate further into the atmosphere. Our results show that these parameters influence WL enhancements, but also suggest that penetration depth alone is not the main moderator of excess WL emission.\\
\indent We find a linear relationship between the Balmer (and Paschen) ratio and the 6684~\AA~continuum enhancements. We note, however, that there are a few cases that do not follow this trend. There is also a very close correlation between the Balmer and Paschen ratios themselves, suggesting a similar formation mechanism. We identify this formation mechanism as (predominantly) hydrogen recombination in optically thin layers, with additional contributions from increased H$^-$ emission as a result of a photosphere heated through radiative backwarming. The extent to which excess H$^-$ emission plays a role varies from case to case. It becomes significant during the declining phase of hydrogen recombination radiation in both of our case studies.\\
\indent There is a large discrepancy between computed values in this work and observed values in the literature. We conclude that purely electron beam-driven situations with the parameters explored in the grid are not realistic for explaining white-light enhancements in solar flares. Special emphasis is put on our inability to explain type II WL flares, since even in cases where electron beam heating creates the right conditions for radiative backwarming, we still see a contribution from optically thin hydrogen recombination that is not observed in real type II WL flares. We conclude that one or more additional energy input mechanisms are needed in order to close the gap between simulations and observations.

\begin{acknowledgements}
    The authors thank the anonymous referee for their detailed and constructive feedback. This work has been supported by the Research Council of Norway through its Centers of Excellence scheme, project number 262622. Computational resources have been provided by UNINETT Sigma2, the National Infrastructure for High-Performance Computing and Data Storage in Norway.
\end{acknowledgements}

\bibliographystyle{aa}
\bibliography{paper_bib}

\end{document}